\documentstyle[prc,twocolumn,aps,epsf,eqsecnum,ifthen,bbox]{revtex}

\newcommand{\beq}{\begin{equation}}
\newcommand{\eeq}{\end{equation}}
\newcommand{\bea}{\begin{eqnarray}}
\newcommand{\eea}{\end{eqnarray}}

\newcommand{\MeV}{\mbox{MeV}}
\newcommand{\GeV}{\mbox{GeV}}

\newcommand{\bl}{\beta_l}
\newcommand{\bp}{\beta_\perp}

\newcommand{\la}{\langle}       
\newcommand{\ra}{\rangle}

\newcommand{\bessk}[1]{{\rm K}_{#1}\! }
\newcommand{\Rs}{R_{s}}         \newcommand{\Rsd}{R_{s}^2}
\newcommand{\Ro}{R_{o}}         \newcommand{\Rod}{R_{o}^2}
\newcommand{\Rl}{R_{l}}         \newcommand{\Rld}{R_{l}^2}
                                \newcommand{\Rold}{R_{ol}^2}

\newcommand{\vyk}{v_{_{\rm YK}}}

\newcommand{\prevc}[3]{Phys. Rev. C {\bf #1}, #3 (#2)}
\newcommand{\prevd}[3]{Phys. Rev. D {\bf #1}, #3 (#2)}
\newcommand{\prevl}[3]{Phys. Rev. Lett.\ {\bf #1}, #3 (#2)}
\newcommand{\zpc}[3]{Z. Phys. C {\bf #1}, #3 (#2)}
\newcommand{\epjc}[3]{Eur. Phys. J. C {\bf #1}, #3 (#2)}
\newcommand{\plb}[3]{Phys. Lett. B {\bf #1}, #3 (#2)}
\newcommand{\npa}[3]{Nucl. Phys. {\bf A#1}, #3 (#2)}
\newcommand{\npb}[3]{Nucl. Phys. {\bf B#1}, #3 (#2)}
\newcommand{\aps}[3]{Acta Phys. Slov. {\bf #1}, #3 (#2)}

\newcommand{\hip}[3]{Heavy Ion Physics {\bf #1}, #3 (#2)}

\newcommand{\NA}[1]{%
    \ifthenelse{\equal{#1}{22}}{EHS/NA22 Coll.}{NA#1 Coll.}}

\begin{document}


\voffset1.5cm


\preprint{TPR-99-12\\CERN-TH/99-215\\nucl-th/9907...}

\title{Reconstructing the Freeze-out State in Pb+Pb Collisions at
        158 $A$GeV/$c$}
\author{Boris Tom\'a\v{s}ik$^a$\thanks{Address after Oct.~1, 1999: 
Department of Physics, University of Virginia, Charlottesville,
Virginia 22901, USA.}, Urs Achim Wiedemann$^{b}$, and 
        Ulrich Heinz$^c$\thanks{On leave of absence from 
        Institut f\"ur Theoretische Physik, Universit\"at Regensburg. 
        E-mail: ulrich.heinz@cern.ch}
}
\address{$^a$Institut f\"ur Theoretische Physik, Universit\"at Regensburg, 
             D-93040 Regensburg, Germany\\
        $^b$Physics Department, Columbia University, New York, NY 10027, USA\\
        $^c$Theoretical Physics Division, CERN, CH-1211 Geneva 23, Switzerland}
\date{\today}

\maketitle

\begin{abstract}
For a class of analytical parametrizations of the freeze-out state
of relativistic heavy ion collisions, we perform a simultaneous 
analysis of the single-particle $m_\perp$-spectra and two-particle 
Bose-Einstein correlations measured in central Pb+Pb collisions 
at the CERN SPS. The analysis includes a full model parameter
scan with $\chi^2$ confidence levels. A comparison of different 
transverse density profiles for the particle emission region allows 
for a quantitative discussion of possible model dependencies of the 
results. Our fit results suggest a low thermal freeze-out 
temperature $T \approx 95 \pm 15\, \MeV$ and a large average transverse 
flow velocity $\bar v_\perp \approx 0.55 \pm 0.07$. Moreover, the fit
favours a box-shaped transverse density profile over a Gaussian one.
We discuss the origins and the consequences of these results in
detail. In order to reproduce the measured pion multiplicity our
model requires a positive pion chemical potential. A study of the 
pion phase-space density indicates $\mu_\pi \approx 60\, \MeV$ for 
$T = 100\, \MeV$. 
\end{abstract} 


\section{Introduction}
\label{intro}

The space-time analysis of hadronic one- and two-particle spectra
measured in relativistic heavy ion collisions has attracted much attention
in recent years. Via the reconstruction of the hadronic phase-space
distribution at freeze-out, this method can provide detailed geometrical
and dynamical information about the last stage of the collision. 
The ultimate goal of the experimental relativistic heavy ion program
is to produce and test the dense early stage of the collision in
which quarks and gluons are the relevant degrees of freedom and a
quark-gluon-plasma is expected (for an up-to-date  overview see 
\cite{QM97,QM99}). However, only the hadronized remnants of this
state are experimentally accessible. Characterizing their spatial
and dynamical distribution, a particle interferometry based space-time
analysis can provide estimates of the phase-space density attained
in the collision and it can establish an experimentally determined
endpoint for microscopic simulations of the complicated multiparticle
dynamics. This makes it a valuable tool in the search for the 
quark-gluon-plasma.

One of the main motivations for space-time analyses in recent years
is the discovery that HBT (Hanbury Brown and Twiss) particle 
interferometry allows to disentangle between random and directed 
dynamical components in the collision 
\cite{CL95,CL96a,Nix-fit,Uli-Hir,WTH97,Rol97}. 
Specifically, in the context of hydrodynamic parametrizations
which provide a very convenient characterization of the freeze-out
region, the one-particle spectra are determined by an effective
blue-shifted temperature from which temperature and flow effects 
cannot be separated unambiguously \cite{SSH93}. On the other hand, 
the $M_\perp$-dependence of the transverse correlation radii 
increases with transverse flow but decreases with stronger thermal 
smearing \cite{MS88}. By combining both observables the ambiguity 
between temperature and flow can be removed.

Knowledge of the magnitude of the transverse flow is crucial for a
dynamical picture of the transverse expansion of the collision 
system and for a dynamical back-extrapolation into the hot and
dense early stage of the collision. With this motivation, there have
been several recent discussions about disentangling temperature
and flow effects. Except for the recent analysis in \cite{App98} most
of  these discussions are published in conference 
proceedings and reviews \cite{WTH97,Rol97,Urs-hab,HJ99,Ster},
and they are mainly based on preliminary data. None of them provides a full
model parameter scan with $\chi^2$ confidence levels for the 
two-particle spectra, and none of them makes quantitative
statements about the model-dependence of the conclusions reached.
However, both these points are very important, since one may wonder,
e.g., to what extent the main conclusions, a relatively low temperature
and high transverse flow velocity, are subject to model-dependent 
details of the parametrization. The present work addresses this gap 
in the literature with an extensive model study. 

Three collaborations measured and published data on Bose-Einstein 
correlations for 158 $A$GeV/$c$ Pb+Pb collisions at the CERN SPS:
WA98 \cite{WA9X}, NA44 \cite{Bea98} and NA49 \cite{App98}. Due to 
their small acceptance, NA44 can only determine particle correlations 
in two transverse momentum bins which moreover correspond to slightly 
different rapidities. Such data are not well suited for a detailed 
ana\-ly\-sis where the $M_\perp$-dependence of the correlation radii plays 
a crucial role. The situation is somewhat better for the recently 
presented WA98 data~\cite{WA9X} but these were not yet available at 
the time of the present analysis. Most suited for our analysis tool 
is a large acceptance experiment like NA49 which covers almost the 
whole forward rapidity region. Their published data are, however, given 
only in the YKP parametrization \cite{App98}. Although in many situations 
its parameters have the most straightforward physical interpretation 
\cite{YKPlett}, this parametrization can be ill-defined in some 
kinematic regions \cite{TH99,diss}, and a cross-check with Cartesian 
(Pratt-Bertsch) correlation radii is hence necessary. The NA49 
experiment has the unique capability of cross-checking experimental 
results from different detector components with overlapping 
acceptance. At the moment, the remaining differences between the 
different detectors are still larger than the published
error bars \cite{Gan-Cat}. We address these subtleties in 
Section~\ref{data} where we discuss in detail the data used in
our analysis. 

The class of fireball models used for our analysis was described 
in detail elsewhere (for recent reviews see e.g. \cite{Urs-hab,HJ99}). 
To get an idea of the degree of model dependence of our final 
conclusions we here investigate two different transverse density 
distributions, a box-shaped and a Gaussian profile. The relevant
model parameters are shortly introduced in Section~\ref{model}. Up 
to now our analysis is restricted to the transverse momentum 
dependence in the particular rapidity bin ($3.9 < Y < 4.4$) for which
a complete set of HBT parameters was published in \cite{App98}.
A realistic description of the rapidity dependence of the transverse 
one-particle spectra would require a refined model 
\cite{Nix-fit,DSH99,MMNN}. The results of the fit, as well
as related technical details, are given in Section \ref{fit}.

As an application of these results we discuss in Section~\ref{psd} 
the average pion phase-space density at freeze-out. Bertsch 
\cite{Ber94} pointed out that knowledge of the single-particle 
spectrum and Bose-Einstein correlations allows for a 
{\em model-independent} extraction of this quantity. The reason 
is that absolutely normalized single-particle spectra carry 
information about the particle density in momentum space while 
the width of the distribution in configuration space can be extracted 
from correlation studies. We extract this model-independent quantity 
from the data and compare it with the prediction of our model; in this 
way we determine the value of the pion chemical potential needed to 
reproduce the measured pion yields. This is an interesting quantity 
since a large value of $\mu_\pi$ would indicate an ``overpopulation'' 
of phase-space and could indicate the onset of multiparticle effects 
(e.g. stimulated emission or a pion laser). 

Section \ref{summary} contains our conclusions. Technical details are
deferred to three Appendices.

\section{The data}
\label{data}

Data on the Bose-Einstein correlation radii as functions of $K_\perp$ 
from the 5\% most central Pb+Pb collisions at 158~$A$~GeV/$c$ were 
published by the NA49 collaboration \cite{App98} for the rapidity 
window $Y\in (3.9,\, 4.4)$ (i.e.\ $1\le Y_{_{\rm CM}} \le 1.5$). Our 
analysis will focus on these data.

The single-particle $p_\perp$-distributions of negatively char\-ged 
hadrons ($h^-$) are taken from \cite{Jones}. At the time of our 
analysis, spectra of identified pions were not yet available for 
the above rapidity window. The differences between $h^-$ and 
identified pion spectra (from negative kaons and a few antiprotons) 
are, however, small since most negatively charged hadrons are pions. 
This is good since these differences will have to be modelled and 
thus introduce (small) systematic uncertainties. The rapidity binning 
of the $h^-$ data in \cite{Jones} is slightly different from that of
the correlation analysis of \cite{App98}. For $h^-$ we use the bin 
$4.15<y<4.65$ which has the largest overlap with that of the correlation 
data. This is an excellent approximation since the $p_\perp$-spectra 
vary only weakly with rapidity in this region \cite{Jones}.

For wide bins the question arises where exactly the data points should 
be placed when comparing them to model predictions. Usually one puts 
them in the middle of the bins. This may, however, lead to systematic 
errors \cite{LW95} if the measured distribution is not flat. Our 
procedure ``where to stick the data point'' is state of the art 
\cite{LW95}: if the measured distribution is well parameterized by a 
function $g(x)$, then the appropriate position of the data point 
corresponding to a bin of width $\Delta x$ between $x_1$ and $x_2$ is 
obtained as
 \beq
 \label{dan1}
   x^{\rm bin} = g^{-1} \left (\frac{1}{\Delta x}\int_{x_1}^{x_2} g(x)\, dx
                        \right ) \, ,
 \eeq
where $g^{-1}$ is the inverse function of $g$. The position of the 
data points of the single-particle spectrum in rapidity is calculated 
from this equation using for $g(x)$ the distribution (\ref{pd1}) 
with $\Delta y = 1.4$ \cite{NA49stop,Gunther}; to get the appropriate 
positions in transverse momentum, the parametrization (\ref{pd5}) with 
$T_{\rm inv} = 185\, \MeV$ (inferred from the measured average transverse 
momentum and the fact that the observed $p_\perp$-spectrum is very well
parameterized by an exponential with inverse slope $T_{\rm inv}$) 
is used for $g(x)$. Note that the width of the $p_\perp$ bins in the 
used data was 100~MeV/$c$, with the first bin starting at 50~MeV/$c$. 
For these data we have explicitly checked that taking the data points 
in the centres of the bins or according to the above described procedure 
does not lead to an observable difference in the fit results.

The HBT two-particle correlation radii published in \cite{App98} for 
the rapidity window $1 \le Y_{_{\rm CM}} \le 1.5$ have several 
shortcomings which can only be resolved in future experimental
analyses: First, the correlation radii are only given for the YKP 
parametrization. Their statistical uncertainties are systematically 
larger than for the Cartesian parametrization \cite{Stefan,Harry}, 
and the data shown in \cite{App98} do not allow for the important 
cross-check \cite{YKPlett,TH99,diss} with the Cartesian parameterization. 
This is an important issue since the YKP fit was recently found to be 
problematic \cite{TH99,diss}. Second, Ref.~\cite{App98} contains only 
data taken in the Main Time Projection Chamber (MTPC), in spite of 
the observed systematic differences between radii extracted from the 
MTPC \cite{Stefan} and the VTPC data \cite{Harry} (of the order of 
0.5~fm \cite{Gan-Cat}). The error bars in \cite{App98} do not 
include this systematic uncertainty which is only roughly estimated 
to be about 15\%. 

To address these difficulties we based our analysis on the three
accessible, but so far unpublished complete data sets from the NA49 
experiment: $h^+h^+$ and $h^-h^-$ correlations from the VTPC 
\cite{Harry} and $h^-h^-$ correlations from the MTPC \cite{Stefan}. 
The reported errors are the output of the MINUIT fitting routine 
which is known to underestimate errors \cite{Sey-priv}. For our 
analysis we considered these three data sets as independent 
measurements and took their average, with correspondingly increased 
error bars which now also include the systematic deviations between 
these measurement. Since both in theory and experiment the 
{\em squared} correlation radii and the YK velocity are the 
directly determined fit parameters \cite{YKPlett,YKPlong}, we 
average over and fit to those.  

\begin{figure}
 \begin{center}
 \epsfxsize=8.5cm
 \centerline{\epsfbox{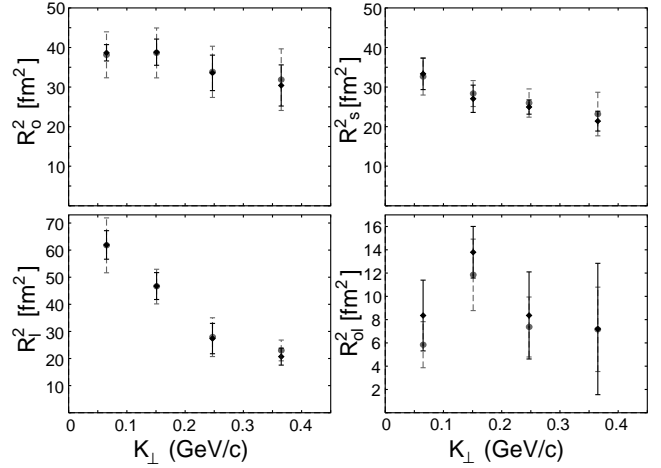}}
\caption{The correlation radii in the Cartesian parametrization 
in the LCMS (see text for details). 
Black symbols indicate the data points from Table~\ref{danal-bpdata}.
Gray dashed symbols represent the values calculated with cross-check
relations from the measured YKP parameters given in the
Table~\ref{danal-ykpdata}.} 
\label{danal-f1}
\end{center}
\end{figure}

\begin{figure}
 \begin{center}
 \epsfxsize=8.5cm
 \centerline{\epsfbox{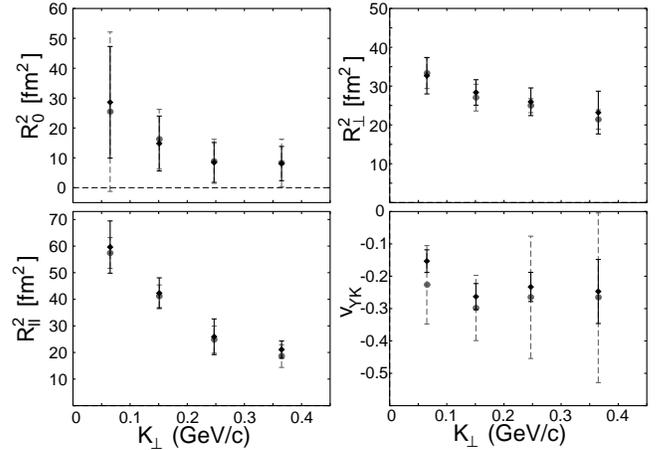}}
\caption{The measured YKP parameters, with $\vyk$ given in the LCMS.
Black symbols indicate the data points from Table~\ref{danal-ykpdata}.
Gray dashed symbols represent the values calculated with cross-check
relations from the measured Cartesian parameters given in
Table~\ref{danal-bpdata}.}  
\label{danal-f2}
\end{center}
\vspace*{-0.75cm}
\end{figure}

The obtained fit parameters for the Cartesian and YKP parametrizations
are summarized in Tables~\ref{danal-bpdata} and \ref{danal-ykpdata}, 
respectively. They are also displayed in Figs.~\ref{danal-f1} and 
\ref{danal-f2}. The minor difference between the rapidity bins 
of the 
VTPC analysis \cite{Harry} ($3.9 < Y < 4.4$) and the MTPC
analysis \cite{Stefan} ($4<Y<4.5$) is irrelevant and has been
neglected. The binning in transverse momentum differs slightly, too. 
Only four of the total number of five $K_\perp$ bins could be taken 
into account in the averaging procedure, and for the fourth bin a slight 
size difference between \cite{Stefan} and \cite{Harry} 
(see Table~\ref{danal-bpdata}) had to be neglected. To position the 
data points in the bins we assumed here that 
their distribution $g(x)$ inside the bin can be linearized. 
Then their position in the bin $[x_1,x_2]$ can be calculated from
\cite{LW95}  
 \beq
 \label{dan2}
   x^{\rm bin} = \frac{\int_{x_1}^{x_2} x\, g(x)\, dx}
                      {\int_{x_1}^{x_2} g(x)\, dx} \,.
 \eeq
Inserting for the function $g(x)$ the distribution $\varrho_2^Y(Y)$ from
equation (\ref{pd4}) with $\Delta y = 1.4$ into (\ref{dan2}), the 
position of the data points in the pair rapidity $Y$ is obtained: 
$Y^{\rm bin}_{_{\rm CM}} = 1.22$. The values of $K_\perp^{\rm bin}$ for 
the different $K_\perp$-bins are calculated using for $g(x)$ the 
function $\varrho_2^{\perp}(K_\perp)$ from equation (\ref{pd13}),
again with $T_{\rm inv} = 185\, \MeV$. The resulting values are 
displayed in Tables~\ref{danal-bpdata} and \ref{danal-ykpdata}. We 
observed \cite{diss} that slightly different (worse) model fits 
are obtained if the data points are placed at the bin centres rather 
than at the positions calculated from (\ref{dan2}). 

It remains to check the compatibility of the Cartesian and YKP
correlation parameters in order to ascertain the validity of the YKP 
data. To this end the Cartesian radii were calculated from the YKP 
parameters and vice versa via cross-check relations published in 
\cite{YKPlett,TH99}. The resulting calculated radius parameters are 
shown in Figs.~\ref{danal-f1} and \ref{danal-f2} by gray dashed 
symbols. The errors were propagated to the calculated radii via 
 \beq
 \label{dan3}
   \sigma_i = \sqrt{\sum_j 
   \left ( \frac{\partial R_i^2}{\partial R_j^{\prime2}} 
   \right )^2 \, \sigma_j^{\prime 2} }\, ,
 \eeq
where $\sigma_i$ is the calculated error of the Cartesian (YKP) 
correlation parameter $R_i^2$ and $\sigma_j^\prime$ the ``measured''
error of the YKP (Cartesian) parameter $R_j^{\prime2}$. (Here the notion
`YKP correlation parameter' includes radius parameters as well as the
YK velocity $\vyk$.) Note that the non-diagonal terms of the error 
matrix are not known and had to be neglected. In most cases the 
calculated parameters show larger errors than the directly measured 
ones. This might improve if non-diagonal error matrix elements could
be taken into account in Eq.~(\ref{dan3}). 

Figs.~\ref{danal-f1} and \ref{danal-f2} show that the YKP and 
Cartesian data are ``on average'' consistent \cite{KK}. In 
view of the known fragility of YKP fits this is an important and
non-trivial result. However, it was recently argued \cite{TH99,diss} 
that the YKP parametrization is quite subtle and can become 
ill-defined for certain (even realistic) sources. The only way to 
test this possibility in experiment is to calculate the YKP 
parameters from the Cartesian ones and check that the resulting 
values are real. Since experimental Cartesian radius parameters have 
a finite measurement error, it is not sufficient to perform this 
consistency check only for their average values; rather, all parameter 
values within the error interval should be checked. When doing so 
we found problems with the definition of the YKP parameters inside 
the error intervals for the third and fourth $K_\perp$ bins. These 
may be related to our further observation \cite{diss} that no good 
model fit was possible starting from the YKP parameters, and that 
the fits tended to drift into strange parameter regions. This 
emphasizes the future need for an explicit cross-check between 
Cartesian and YKP parameter fits directly on the experimental 
level, including error propagation with the complete error matrix. 
Due to our problems with the measured YKP parameters, the model fits 
presented here are based on an analysis of the Cartesian correlation 
radii. 

\section{The model}
\label{model}

\subsection{The emission function}
\label{emfun}

Our model analysis is based on the widely used emission function
\cite{CL95,CL96a,YKPlong,WSH96,TH98,CSH95b} 
\bea
  \nonumber
  S(x,K)\, d^4 x & = & \frac{M_\perp \cosh (Y - \eta)}{(2\pi)^3}\, 
  \exp\left ( - \frac{K \cdot u(x)}{T} + \frac{\mu}{T} \right )\\ 
  && \nonumber \times G(r)\, 
  \exp \left ( - \frac{(\eta - \eta_0)^2}{2 (\Delta\eta)^2} \right )\, 
  d\eta\, r\, dr\, d\varphi\\
  && \times
  \frac{\tau\, d \tau}{\sqrt{2 \pi (\Delta\tau)^2}} \,
  \exp \left ( - \frac{(\tau - \tau_0)^2}{2 (\Delta\tau)^2} \right )\, .
  \label{mod1}
\eea
Here, the pair momentum $K$ is parameterized in terms of the transverse mass 
$M_\perp = \sqrt{K_\perp^2 + m^2}$, longitudinal
rapidity $Y$, and the transverse momentum $K_\perp$ 
\beq
  \label{mod2}
  K = (M_\perp \cosh Y,\, K_\perp,\, 0,\, M_\perp \sinh Y)\, .
\eeq
We use here the usual coordinate system with x-axis directed parallel to 
$\bbox{K}_\perp$. In configuration space, we use the polar coordinates $r$ 
and $\varphi$ for the transverse directions, a longitudinal proper time
$\tau=\sqrt{t^2 - z^2}$ and space-time rapidity $\eta =
\frac12\ln\frac{t+z}{t-z}$. 

The $\eta$-profile in the model is chosen to be a Gaussian peaked at
$\eta_0$ with width $\Delta\eta$. In this way $\eta_0$ determines
the centre of mass (CMS) rapidity. Since we focus on a single rapidity 
bin and work in the LCMS, we set $\eta_0 = -1.22$ (see previous Section) and
$\Delta\eta = 1.3$. The latter value is obtained from a comparison with 
the single-particle rapidity distribution. 

Freeze-out proper times are distributed according to a Gaussian of
width $\Delta\tau$ centred at $\tau_0$. This allows for freeze-out of
particles during an extended time period given by $\Delta\tau$. 

The transverse geometry is specified by the distribution 
$G(r)$ in (\ref{mod1}). As only azimuthally symmetric fireballs are
considered, this function does not depend on $\varphi$. In our
analysis two particular profiles will be assumed: a {\em Gaussian} one
\beq
  \label{mod6}
  G(r) = \exp \left ( - \frac{r^2}{2R_G^2} \right )\, ,
\eeq
and a {\em box-shaped} one
\beq
  \label{mod7}
  G(r) = \theta (R_B - r)\, .
\eeq

The fireball at freeze-out is assumed to be in local thermal equilibrium 
at temperature $T$. This is modelled by the Boltzmann factor in 
(\ref{mod1}). The chemical potential $\mu$ in its argument has no 
effect on the correlation radii and only affects the normalization 
of the single particle spectrum. The argument $K{\cdot}u(x)$ allows 
for longitudinal and transverse expansion of the fireball through 
the collective four-velocity field $u(x)$:
\bea
  u(x) &=& \left(\cosh\eta_t\, \cosh\eta, \,\cos\varphi\, \sinh\eta_t, \,
           \sin\varphi\, \sinh\eta_t, \right.
           \nonumber \\
           && \left. \qquad \qquad \qquad \cosh\eta_t\, \sinh\eta\right)\, .
  \label{mod8}
\eea
The longitudinal expansion rapidity $\eta_l(x)$ has here been identified
with the space-time rapidity $\eta$. This leads to a longitudinal 
expansion velocity $v_{\rm long}(z,t) = \tanh \eta = z/t$, corresponding 
to Bjorken-type boost-invariant expansion \cite{Bjork}. The transverse 
expansion is parameterized by the transverse flow rapidity $\eta_t(x)$. 
It is assumed to increase linearly with the distance from the collision 
axis,
\beq
  \label{mod9}
  \eta_t(x) = \eta_f \frac{r}{r_{\rm rms}}\, .
\eeq
The scaling factor $\eta_f$ specifies the value of the transverse 
flow rapidity at the transverse rms radius, given by
\beq
  \label{mod10}
  r_{\rm rms} = \sqrt{2} R_G
\eeq
for the Gaussian transverse distribution and by
\beq
  \label{mod11}
  r_{\rm rms} = \frac{R_B}{\sqrt{2}}
\eeq 
for the box-shaped one. From these relations we expect the two fit 
parameters in (\ref{mod6}) and (\ref{mod7}) to satisfy approximately 
$R_B \approx 2\, R_G$.

In the literature the transverse flow is often quoted in terms of the 
{\em average transverse expansion velocity} $\bar v_\perp$
\beq
  \label{mod-meanvel}
  \bar v_\perp = \frac{\int_0^\infty r\, dr\, \tanh\eta_t(r) \, G(r)}
                      {\int_0^\infty r\, dr\, G(r)} \, .
\eeq
Its value is roughly given by the radial velocity at the transverse 
rms radius, $v_\perp (r_{\rm rms}) \simeq \tanh\eta_f$; looking more 
closely, it is slightly smaller: for $\eta_f$=0.6 one finds 
$v_\perp(r_{\rm rms}){\approx}0.54$ and $\bar v_\perp$=0.50 (0.46) for the 
box-like (Gaussian) transverse density profile.    

Our calculation of the single-particle spectra from (\ref{mod1}) 
includes contributions from resonance decays as described in 
\cite{SSH93,WH97res}. This is crucial for a correct description 
of the yield and shape of the spectra. We include all decays with 
branching ratios above 1\%  from mesons with masses up to 1020 MeV/$c^2$ 
and from baryons with masses up to 1400 MeV/$c^2$. For decay chains,
the pro\-duct of the branching ratios is required to be larger than 1\%.
For baryons the chemical potential $\mu_B(T)$ was parameterized 
according to (\ref{chempot}). Strange particles acquire a chemical 
potential $\mu_S(T)$ determined by the condition (\ref{stpot}) of 
strangeness neutrality. Finally, contributions from negative kaons 
and antiprotons were included in the calculation of the $h^-$ spectrum.

The role of resonance decay contributions to the HBT radius parameters 
is known to be much less important \cite{WH97res}. Our calculation of 
the correlation radii will thus only include direct pions. This is 
an essential technical simplification: the additional integrals from 
the resonance decay phase space would have increased the numerical 
task from calculating a 2-dimensional integral to calculating a  
5-dimensional (for two-particle decays) or 6-dimensional one (for 
three-particle decays). This very time consuming calculation was
performed in model studies \cite{WH97res} where the correlator was 
evaluated for only a few characteristic sets of model parameters. 
However, in a simultaneous multi-parameter fit, in which the fit
routine calls the two-particle correlator approximately 50-100 times 
for thousands of different model parameter combinations, it cannot be
done.

We finally remark that the model (\ref{mod1}) is customarily used
with a Boltzmann distribution rather than a Bose-Einstein one, since
this greatly simplifies the analytical and numerical analysis. The
resulting differences are usually small. If the pions develop a 
positive chemical potential, however, this visibly affects the shape 
of the single particle spectra; for this reason we use, in fact, 
a Bose-Einstein distribution to calculate the pion multiplicities 
and the single-particle spectra of direct pions (see Appendix 
\ref{comps} and \ref{mult}). For the heavier resonances the Boltzmann 
approximation is excellent. For technical reasons we also use the
Boltzmann approximation in the calculation of the correlation radii, 
in line with the much larger systematic error of the corresponding 
available data.

\subsection{Basic relations}
\label{basrel}

We shortly recall how the emission function $S(x,K)$ is related
with the observables to be calculated. More details can be found e.g. 
in \cite{Urs-hab}.

The single-particle spectrum is obtained from
\beq
  \label{rel1}
  E_p\frac{dN}{d^3p} = P_1(p) = \int d^4x\, S(x,p)\, .
\eeq
For the two-particle correlations, we used the Cartesian
parametrization:
 \bea
 \label{rel2}
   C(q,K) - 1 
   &=& \exp\left[- q_o^2\Rod(K) - q_s^2 \Rsd(K) \right. 
 \nonumber \\
   && \qquad \left. - q_l^2 \Rld(K) - 2q_oq_l \Rold(K) \right]\, ,
 \eea
where the $q_i$ are the components of the momentum difference in the
{\em out-side-long} coordinate system and $K$ stands for the average
pair momentum. The {\em correlation radii} are obtained from the
emission function via
\begin{mathletters}
 \label{rel3}
 \bea
   \Rsd & = & \la \tilde y^2 \ra \, , \label{Rside} \\
   \Rod & = & \la (\tilde x - \bp \tilde t)^2 \ra \, , \label{Rout} \\
   \Rld & = & \la ( \tilde z - \bl \tilde t )^2\ra \, ,  \label{Rlong} \\  
   \Rold & = & \la (\tilde x - \bp \tilde t)(\tilde z - \bl \tilde t)\ra
   \, .
 \label{Routlong}
 \eea
\end{mathletters}
Here 
 \beq 
 \label{rel4}
   \tilde x_\mu = x_\mu - \bar x_\mu \, , \qquad 
   \bar x_\mu(K) = \langle x_\mu \rangle \, ,
 \eeq
and
 \beq
 \label{meas11}
   \langle f(x) \rangle(K) = \frac{\int d^4x\, f(x)\, S(x,K)}
                                  {\int d^4x\ S(x,K)}\, .
 \eeq

In our calculations all integrations were performed numerically. This 
distinguishes our work from Ref.~\cite{Ster} where analytic 
approximations were used instead. For sources with strong transverse 
flow (as will be the case here) the analytical approximations for 
the correlation radii may become problematic \cite{WSH96,TH98,CLL96}. 
Also, it is not easy to capture the intricate effects of resonance 
decay kinematics on the single-particle spectra with simple 
ana\-ly\-tical expressions as those used in \cite{Ster}.  

\section{The Fit}
\label{fit}

\subsection{Single-particle $p_\perp$-spectrum}
\label{sinspec}

The first step is to fit the $h^-$ single-particle $p_\perp$-spectrum
with the model (\ref{mod1}). Resonance decay contributions are treated 
as described in Sec.~\ref{emfun}. The geometrical model parameters 
$\Delta\eta$, $R_B$ ($R_G$), $\tau_0$, $\Delta\tau$ are known to only 
affect the absolute normalization of the single-particle spectrum, but 
not its shape \cite{SSH93,Urs-hab}. For fixed transverse density and 
flow profiles, the spectral shape is completely determined by the 
temperature $T$ and the transverse flow strength $\eta_f$. 

The fit to the spectrum was performed with the CERN package MINUIT 
\cite{minuit}, using for the calculation of (\ref{rel1}) (see 
Appendix~\ref{comps}) a routine described in \cite{WH97res}. The 
data were compared with the $y$-integrated $m_\perp$-spectrum 
rather than with the spectrum at a fixed value of $y$. Since the
$y$-integration can be done analytically and removes the dependence
on the longitudinal kinematic limits for resonance decays, the 
former requires much less computation time. The differences are very 
small since the $m_\perp$-spectra obtained from (\ref{mod1}) are almost 
$y$-independent, except for very forward/backward rapidities. 

\begin{figure}
\begin{center}
\begin{minipage}{7.1cm}
 \epsfxsize=6.9cm
 \centerline{\epsfbox{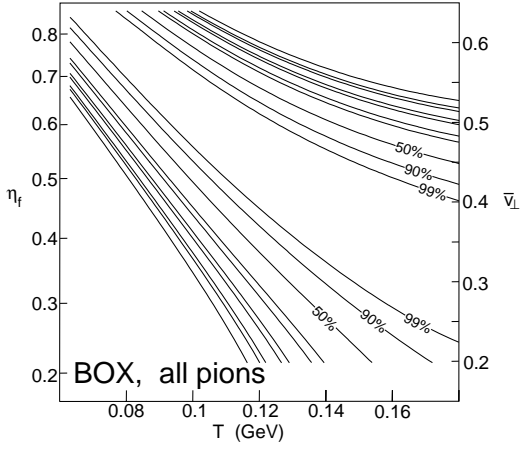}}
 \centerline{{(a)}}
\end{minipage} 
\begin{minipage}{7.1cm}
 \epsfxsize=6.9cm
 \centerline{\epsfbox{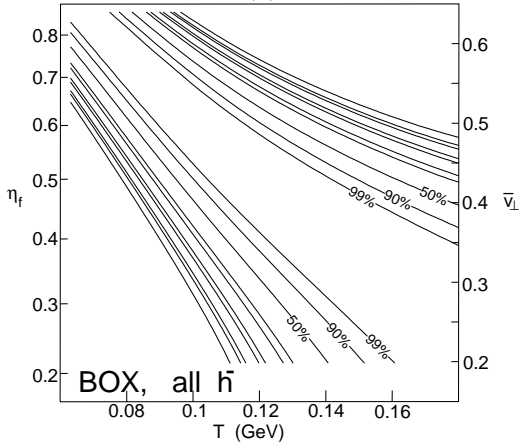}}
 \centerline{{(b)}}
\end{minipage}
 \caption{$\chi^2$ contour plots for the fit of the single-particle $h^-$
 spectrum to the model (\protect\ref{mod1}) with a box-shaped transverse 
 density profile. The contour lines correspond to the following confidence 
 levels: 99\%, 90\%, 50\%, 10\%, 5\%, 1\%, 0.5\%, 0.1\%, 0.05\%, and 0.01\%. 
 Panel (a): fit including all pions. Panel (b): negative hadrons, including
 also $K^-$ and $\bar p$ and their resonance decay contributions.
 \label{danal-f3}}
 \end{center}
\vspace*{-.5cm}
\end{figure}

\begin{figure}
\begin{center}
\begin{minipage}{7.3cm}
 \epsfxsize=7.1cm
 \centerline{\epsfbox{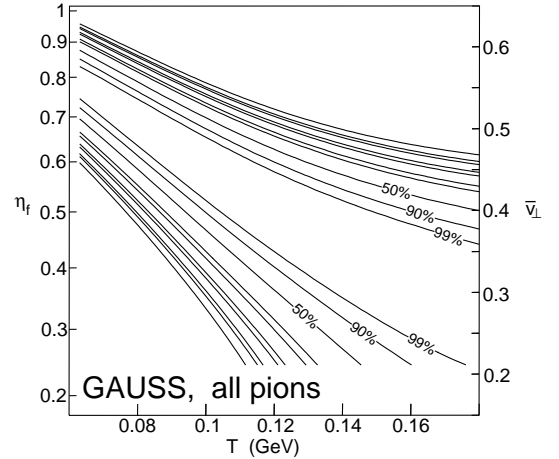}}
 \centerline{{(a)}}
\end{minipage}
\begin{minipage}{7.3cm}
 \epsfxsize=7.1cm
 \centerline{\epsfbox{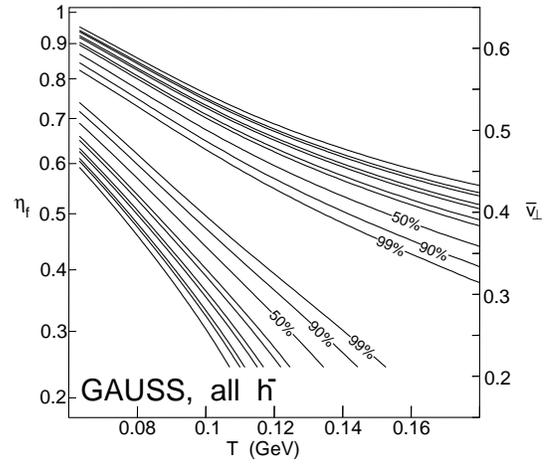}}
 \centerline{{(b)}}
\end{minipage}
 \caption{The $\chi^2$ contour plots as in Fig.\ 
 \ref{danal-f3} but for the model (\protect\ref{mod1}) with 
 Gaussian transverse density profile. 
 The  average transverse expansion  velocity $\bar v_\perp$  
 allows for a direct comparison between the two models. The corresponding
 values of $\eta_f$ are also shown; they differ substantially between the 
 two models.} 
 \label{danal-f4}
 \end{center}
\vspace*{-0.4cm}
\end{figure}

The resulting $\chi^2$ contour plots in $T$ and $\eta_f$ are shown in
Figs.~\ref{danal-f3} (box-shaped transverse geometric profile) and
\ref{danal-f4} (Gaussian profile). For comparison the average transverse
expansion velocity $\bar v_\perp$ is given on the right ordinates of 
these Figures. For their calculation the routine was driven through 
the whole $T$-$\eta_f$ domain covered by the Figures, and in each point 
the normalization of the spectrum was fitted as a third parameter. 
One finds a clear $\chi^2$ ``valley'' pointing from the upper left to 
the lower right corner, reflecting the anticorrelation between $T$ and 
$\eta_f$ in the $m_\perp$-slope. The indicated confidence levels result 
from a $\chi^2$-distribution with 16 degrees of freedom (19 data points 
minus 3 fit parameters) \cite{abram}. 

The two panels of Fig.~\ref{danal-f3} show the effects of inclu\-ding 
the $K^-$ and $\bar p$ contributions in the negative hadron spectra --
the upper panel includes only negative pions (including all resonance
decays). At low temperature the fit results are nearly identical
because the $K^-$ and $\bar p$ contributions are strongly suppressed
by their masses and by the large baryon chemical potential (at $T$=80
MeV we have $\mu_B  \approx 430$\,MeV), but at higher temperatures
their inclusion reduces appreciably the amount of transverse flow
needed to fit the measured slope. Our results differ in two points
from those published in \cite{WTH97}: (i) We include nonzero baryon
and strangeness chemical potentials. This increases the decay
contributions from baryon resonances, making the spectra steeper and
thus requiring more transverse flow to reproduce the measured
slope. Compared to the case of vanishing chemical potentials this
shifts the ``$\chi^2$ valley'' in Fig.~\ref{danal-f3}a upwards by
$\approx$\,0.05\,$c$ (somewhat less at low $T$ and more for higher
$T$). (ii) We include contributions from $K^-$ and $\bar p$. This 
flattens the $h^-$-spectrum further because these heavier particles are more
strongly affected by transverse flow. To fit a given spectral slope,
lower values of $\eta_f$ are thus needed at given $T$. The two effects
(i) and (ii) are seen to partially cancel each other.

A comparison of Figs.~\ref{danal-f3} and \ref{danal-f4} gives a 
feeling for the model-dependence of the fit resulting from two 
different choices for the transverse density profile. Since at 
fixed $\eta_f$ they lead to different average transverse flow 
velocities $\bar v_\perp$ (the quantity which determines the 
blueshift of the spectral slope), one should use the labelling on 
the r.h.s. of these Figures for comparison. One sees that for a 
Gaussian density profile slightly smaller va\-lues for 
$\bar v_\perp$ are needed to account for the measured slope. The 
reason is that superimposing a Gaussian distribution $G(r)$ with a 
monotonically increasing transverse flow profile, a significant 
part of the high-$p_\perp$ part of the spectrum is obtained from 
contributions at large transverse distances $r > R_G$ in the 
Gaussian tails. A box profile does not allow for emission from
distances $r > R_B$ and thus requires slightly larger values for
$\bar v_\perp$ to account for the same spectra. 

Some access to the shape of the transverse density profile is possible 
through two-particle correlations (see Sec.~\ref{correl}), but limited 
statistics and other uncertainties leave some room for interpretation. 
The differences between Figs.~\ref{danal-f3} and \ref{danal-f4} should 
thus be mainly taken as an estimate for the systematic model 
uncertainties in the finally extracted values for $T$ and $\bar
 v_\perp$. That they are small for the single-particle fits is 
certainly gra\-ti\-fying.

\subsection{Two-particle correlations}
\label{correl}

The correlation radii summarized in Table~\ref{danal-bpdata} were 
fitted with the CERN package MINUIT \cite{minuit}, using for the 
model calculation a routine which computes the correlation radii 
from (\ref{rel3}). We scanned the whole $T$-$\eta_f$ domain as before, 
performing in each point a 3-parameter fit to find the best values 
of $R_B$ ($R_G$), $\tau_0$, and $\Delta\tau$. $\Delta\eta=1.3$ was 
always kept fixed, see Sec.~\ref{model}.

\begin{figure}
\begin{center}
 \epsfxsize=7.1cm
 \centerline{\epsfbox{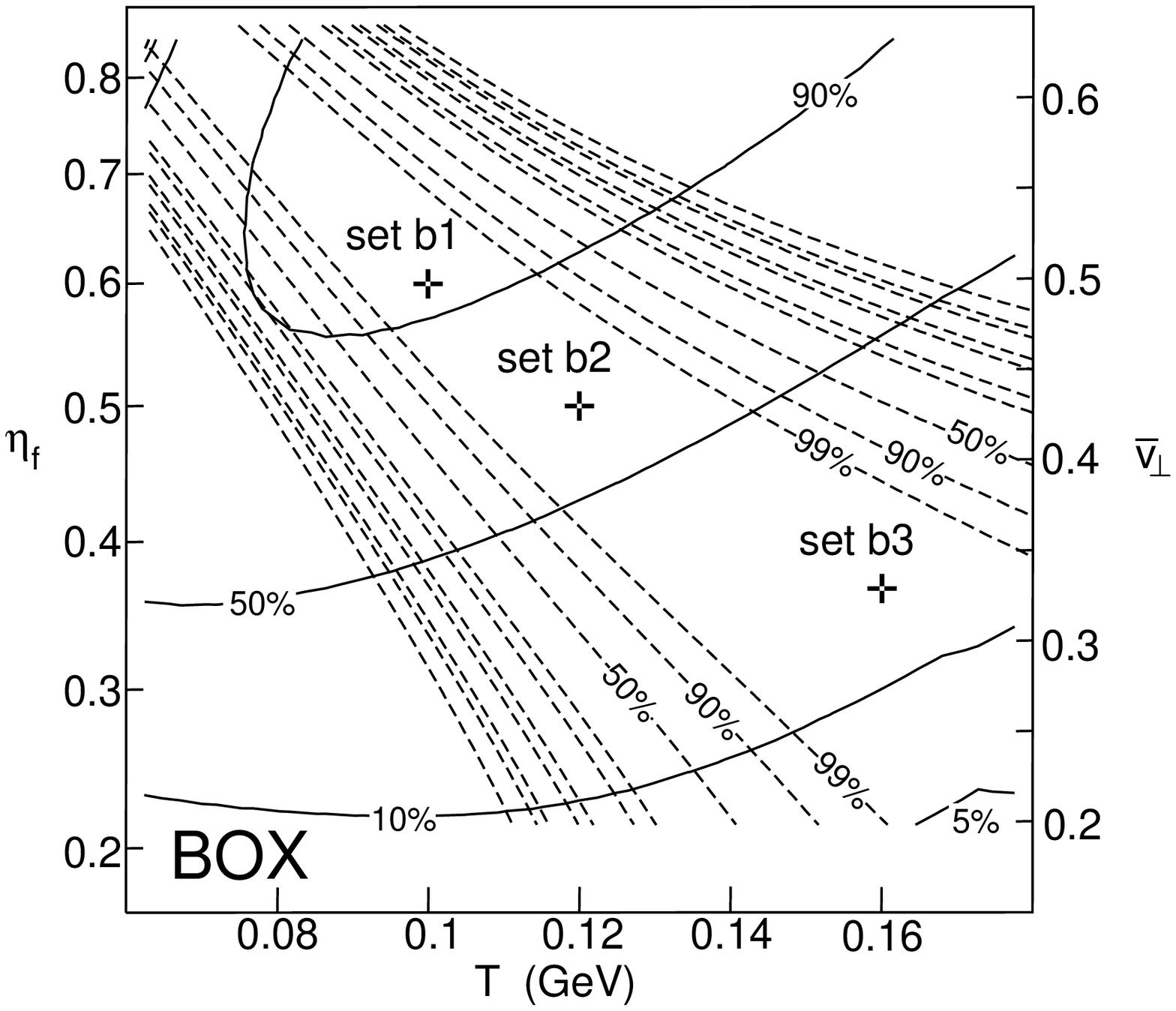}}
 \caption{$\chi^2$ contours for the fit of the Cartesian HBT
 radii to the model (\protect\ref{mod1}) with a box-shaped transverse 
 density profile (solid lines). The numbers give the corresponding 
 confidence levels. The dashed lines show the result of the fit to 
 the single-particle spectrum (Fig.\ \ref{danal-f3}b). Crosses denote 
 the positions of the three parameter sets used in further calculations.} 
 \label{btokyo}
 \end{center}
\vspace*{-0.5cm}
\end{figure}

\begin{figure}
\begin{center}
 \epsfxsize=7.1cm
 \centerline{\epsfbox{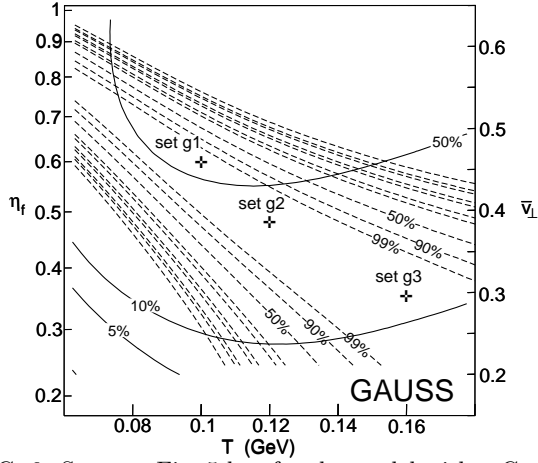}}
 \caption{Same as Fig.~\ref{btokyo} but for the model with a
 Gaussian transverse density profile.} 
 \label{gtokyo}
\end{center}
\vspace*{-0.5cm}
\end{figure}

The results of these fits, superimposed on the fit of the single-particle
spectra, are shown in Figs.~\ref{btokyo} (box-profile) and \ref{gtokyo} 
(Gaussian profile). The contours correspond to a $\chi^2$-distribution 
of 11 degrees of freedom (16 data points minus 5 model parameters).
These results will be further discussed in Sec.~\ref{rd}. Here we only
observe that the HBT radii indeed allow to disentangle the ambiguity 
between temperature and transverse flow (although the uncertainties 
are still significant). The box-shaped transverse geometric profile 
seems to be favoured by the fit, although limited statistics does not 
allow to rule out a Gaussian shape. Independent of the choice of the 
transverse density profile we can, however, safely conclude that
the data require strong transverse flow with $\bar v_\perp > 0.3$.
The best fits favour low freeze-out temperatures between 80 and 110 MeV 
and large average transverse expansion velocities between 0.47\,$c$ to 
0.62\,$c$ (for the box model).

On the other hand, the ``$\chi^2$ valley'' defined by the correlation 
radii is clearly seen to deviate from the simple dependence on 
$\eta_f$ and $T$ which one obtains in analytical approximation 
for Gaussian transverse density profiles \cite{CSH95b,CNH95}:
 \beq
 \label{dan4}
   R_s^2(M_\perp) \approx R_G^2\, 
   \left ( 1 + \frac{\eta_f^2}{2T} M_\perp \right )^{-1}\, .
 \eeq
Analyses employing this relation (e.g.~\cite{Ster,App98}) should 
there\-fore be taken with some caution. 

\subsection{Total yield}
\label{yield}

Our fit determines for each combination ($T,\eta_f$) of temperature 
and transverse flow the remaining model para\-me\-ters. This allows for the
calculation of the total pion multiplicity by integrating the 
model emission function over the whole phase-space. The corresponding 
formulae are derived in Appendix~\ref{mult}. We recall that for this 
calculation we use a Bose-Einstein distribution for the direct pions 
and that resonance decay contributions are included. 

\begin{figure}
\begin{center}
\epsfxsize=8.6cm
\centerline{\epsfbox{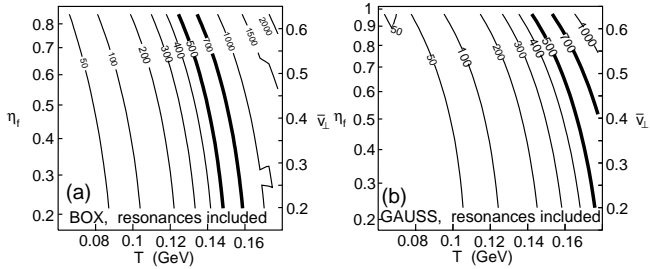}}
\vspace*{0.2cm}
\caption{
Total $\pi^-$ multiplicity (resonance decays included) as a function 
of $T$ and $\bar v_\perp$, computed from the model with (a) box-shaped, 
(b) Gaussian transverse density profile. The wiggles in the contours 
towards the right in panel (a) result from numerical uncertainties in 
the fitted model parameters. For other details see text.
\label{f-mult}}
\end{center}
\vspace*{-1.1cm}
\end{figure}

The resulting $\pi^-$ multiplicities (including pions from resonance 
decays) are shown in Fig.~\ref{f-mult} for both box-like and Gaussian 
transverse density profiles. At very low temperatures (up to 100~MeV) 
the total pion multiplicity is dominated by directly produced pions.
With increasing temperature the fraction of pions from resonance decays 
grows rapidly. For the model with a box-shaped density profile the
multiplicity grows faster with temperature; for $\eta_f=0$ this model 
produces (in Boltzmann approximation) twice as many pions as a Gaussian 
model whith the same transverse rms radius (i.e.\ with $R_G = R_B/2$). 
This is due to the larger covariant volume occupied by the box-shaped
model. 

The experimentally measured negative hadron multiplicity $715\pm 30$ 
\cite{NA49stop,Gunther} is indicated by thick contour lines. Clearly, 
these ``bands'' lie far outside the regions of ($T,\eta_f$) favoured 
by the fit to the spectrum and correlation radii (cf. Figs.~\ref{btokyo} 
and \ref{gtokyo}). They tend to favour much higher temperatures. 
Again, the model with a box-like density profile is favored since
it minimizes this discre\-pancy. Nevertheless, a non-zero pion chemical
potential must be introduced in both models for a correct reproduction 
of the measured multiplicity. This question is studied in more detail in
Sec.~\ref{psd}.

%
\subsection{Results and discussion}
\label{rd}

In order to obtain a more direct picture of the quality of the above 
fits, we selected three sets of fit parameters for each of the two
models (Gaussian ({\sf g1, g2, g3}) and box-shaped ({\sf b1, b2, b3})), 
indicated in Figs.~\ref{btokyo} and \ref{gtokyo}. The corresponding 
complete parameter sets are given in Table~\ref{danal-sets} which also 
shows the predicted total $\pi^-$ multiplicities (including
resonance decays) in each case. 

\begin{figure}
 \begin{center}
 \epsfxsize=6.5cm
 \centerline{\epsfbox{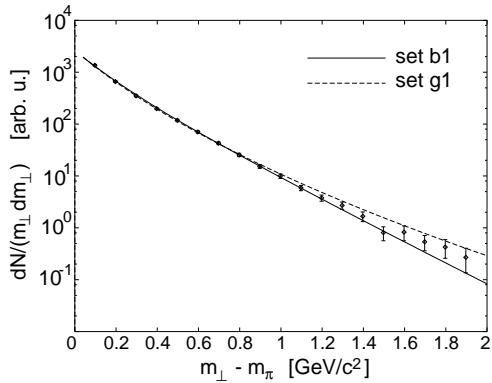}}
\caption{Measured $h^-$ spectrum in the rapidity window $4.15{<}y{<}4.65$
 \protect\cite{Jones} compared to the best model fits {\sf b1} and 
 {\sf g1}.
 \label{specporov}}
\end{center}
\vspace*{-1.1cm}
\end{figure}

\begin{figure}
 \begin{center}
 \epsfxsize=8.5cm
 \centerline{\epsfbox{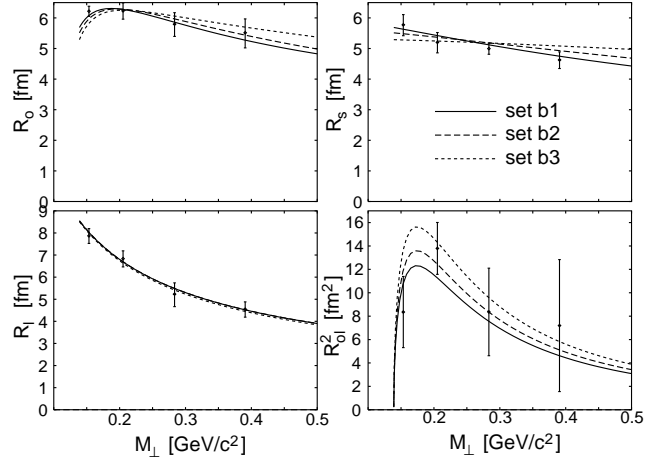}}
\caption{Comparison of model predictions from the model with box-shaped
transverse geometric profile and inserted parameter sets {\sf b1, b2, b3}
with the data on Cartesian correlation radii (to which the model was fitted).}
\label{danal-f14}
\end{center}
\vspace*{-0.8cm}
\end{figure}

\begin{figure}
 \begin{center}
 \epsfxsize=8.5cm
 \centerline{\epsfbox{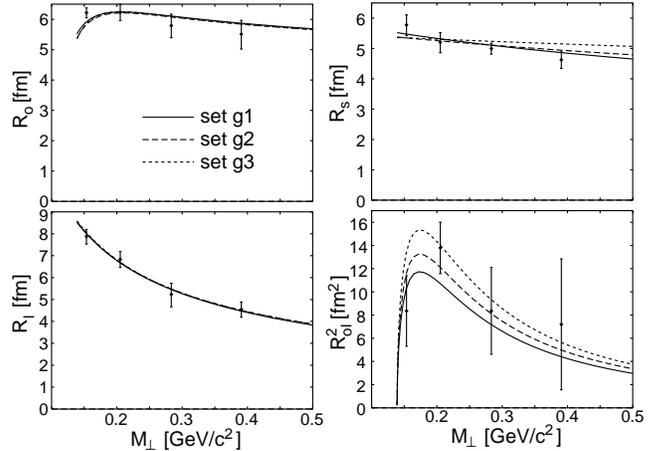}}
\caption{Same as Fig.~\ref{danal-f14} but for the Gaussian transverse 
density profile with parameter sets {\sf g1, g2, g3}.}
\label{danal-f15}
\end{center}
\vspace*{-0.5cm}
\end{figure}

The quality of the fits {\sf b1} and {\sf g1} to the measured 
single-particle $m_\perp$-spectrum \cite{Jones} is seen in 
Fig.~\ref{specporov}. Similarly, Figs.~\ref{danal-f14} (box) and 
\ref{danal-f15} (Gauss) show the quality of the fits to the Cartesian 
correlation radii. The box-shaped density profile accommodates the 
$M_\perp$-dependence of $\Ro$ and $\Rs$ better than the Gaussian one.
It reproduces the rapid initial increase of $\Ro$, its rather steep 
decrease at larger $M_\perp$, and the slope of $\Rs$ all reasonably well. 
With the ($T,\eta_f$) values allowed by the single-particle spectrum, 
the Gaussian model has difficulties in reproducing the strong 
$M_\perp$-dependence in $R_s$ and $R_o$.

\begin{figure}
\begin{center}
\epsfxsize=8cm
\centerline{\epsfbox{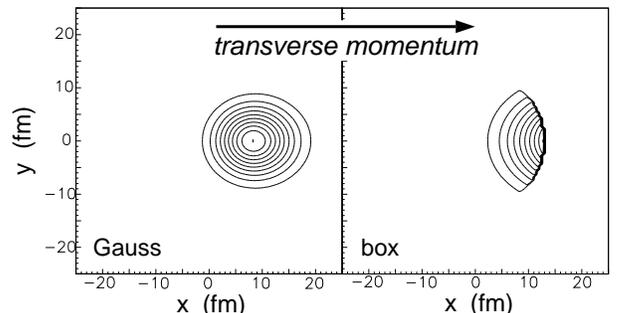}}
\caption{Transverse cuts through the effective emission region
of particles with $K_\perp$=500 MeV/$c$. Left: Gaussian transverse 
density profile. Right: box-like transverse density profile. In both 
cases $T$=100\,MeV, $\eta_f$=0.8.
\label{profs}}
\end{center}
\vspace*{-0.5cm}
\end{figure}

The different fit qualities achieved by these two models reflect 
their different behaviour in the presence of transverse
flow. This is illustrated in Figure \ref{profs}, where 
transverse cuts through the 
effective emission region for particles with $K_\perp$=500 MeV/$c$
(in $x$-direction) are shown. For the Gaussian source
the effective emis\-sion region moves parallel to $\bbox{K}_\perp$ 
outward into the tail of the density profile; its ``outward'' 
homogeneity length is {\em larger} than its ``sideward'' one. For 
the box-shaped di\-stribution, which forbids particle emission 
from $r>R_B$, the opposite is found: the effective emission region gets 
squeezed towards the edge of the box, and the ``outward'' homogeneity 
length is now {\em smaller} than the ``sideward'' one. This causes a 
more rapid decrease of $R_o$ with rising $K_\perp$ in this case, and it 
is this feature of the model which is preferred by the data.

Recent studies of deuteron production via coalescence appear to support 
our conclusion. They also favour box-shaped transverse density 
distributions, although for a different reason \cite{Pol,SH98}:
the box-profile gives more weight to regions of large transverse 
flow velocities, and only in this way can one understand the observed
flattening of the deuteron $m_\perp$-spectra compared to the proton ones
\cite{Pol,SH98}. In contrast, what matters here for the correlation 
radii is the increasingly negative contribution from $\la \tilde x^2
 - \tilde y^2 \ra$ to $\Rod - \Rsd$ at large $M_\perp$. This property 
was originally attributed in \cite{HV98} to opacity effects in the 
source, i.e. to the suppression of particle emission from the interior 
of the source in favor of surface emission. In this sense, a radially 
expanding source with a box-like density profile looks at large 
$K_\perp$ like an ``opaque source'' (see Fig.~\ref{profs}).

This ambiguity illustrates that particle interferometric measurements 
can make statements about the rms widths of the homogeneity regions, 
but that it is a model-dependent task to interpret how these
homogeneity regions and their $\bbox{K}$-dependences are generated. 
That transverse density profiles with a sharper edge than the Gaussian
one can naturally explain both the ``opacity effects'' in the 
correlation radii at large $K_\perp$ and the stronger flow experienced 
by the deuterons may be taken as an indication, but not as proof that 
they form indeed the preferred parametrization of the source at 
freeze-out.

Before turning to a discussion of the remaining two radius parameters, 
$\Rl$ and $\Rold$, let us stress another very important feature 
of the source: its strong {\em transverse growth} before freeze-out.
Both the box-like and Gaussian transverse density profiles give at 
freeze-out transverse rms radii $r_{\rm rms} = R_G\sqrt{2} = R_B/\sqrt{2}$ 
of $\approx$ 9 fm. This is about twice the transverse rms radius of the 
original overlapping Pb-nuclei of $\approx$ 4.5 fm. In view of the 
results obtained in \cite{WH97res} for a Gaussian density profile it 
seems unlikely that the neglected resonance decay contributions can 
account for a significant fraction of this large difference, although 
we did not check this possibility explicitly again for the box-like 
distribution. Dynamical consistency requires that such a strong 
geometric growth is accompanied by strong radial flow, as indeed seen 
in our analysis.

The longitudinal radius $\Rl$ is fitted very well by all selected 
parameter sets. This supports the validity of the assumed Bjorken 
scenario of boost-invariant longitudinal expansion of the reaction 
zone at thermal freeze-out. The measurement of the time parameters 
of the model ($\tau_0$ and $\Delta\tau$) is, however, affected by 
rather large errors, and it is model-dependent. The model-dependence
of $\Delta\tau$ was extensively discussed in \cite{TH99} to which we 
refer for details. Here we concentrate on a discussion of the 
parameter $\tau_0$.

For a system undergoing boost-invariant longitudinal expansion, the
size of $\Rl$ is dominated by the longitudinal flow velocity gradient
\cite{CSH95b} whose inverse grows linearly with the longitudinal proper
time $\tau$. Under these conditions it is suggestive to interpret 
$\tau_0$ (as extracted from $\Rl$) as the total time from impact to 
freeze-out \cite{MS88}. For the sets of fit parameters listed in 
Table~\ref{danal-sets} this interpretation runs, however, into 
trouble: taking the measured final average transverse flow velocity 
of $\bar v_\perp \approx 0.5\,c$ and assuming constant acceleration
one would expect \cite{HSpriv} the mean (rms) radius of the matter to 
expand roughly according to $r_{\rm rms}(\tau) = r_{\rm rms}(0) +
 {1\over 2} \bar v_\perp(\tau)\tau$. (This slightly exaggerates the 
point to be made since the higher pressure will lead initially to 
stronger acceleration.) This expression should be evaluated at the 
average emission time which for the source (\ref{mod1}) is given by 
$\langle\tau\rangle = \tau_0 + {(\Delta\tau)^2\over\tau_0}$. For the 
two preferred parameter sets {\sf b1} and {\sf g1} in 
Table~\ref{danal-sets} we get $\langle\tau\rangle \approx 8.4$\,fm/$c$. 
With $r_{\rm rms}(0)\approx 4.5$\,fm one thus finds 
$r_{\rm rms}(\langle\tau\rangle)\approx 6.6$ fm; this falls clearly 
short of the measured rms radius of $8.5-9$ fm.

\begin{figure}
 \begin{center}
 \epsfxsize=7.5cm
 \centerline{\epsfbox{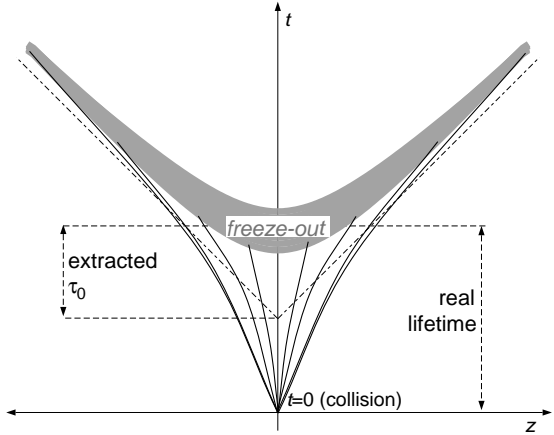}}
\caption{Schematic sketch of the interferometric lifetime
measurement. The time parameter $\tau_0$ is extracted under the
assumption that the longitudinal expansion is boost-invariant 
during the entire dynamical evolution. If the boost-invariant 
velocity profile develops gradually, the fireball can live longer.} 
\label{expan}
\end{center}
\vspace*{-0.5cm}
\end{figure}

For realistic expansion scenarios the time necessary to expand 
to such large radii thus exceeds both $\tau_0$ and
$\langle\tau\rangle$. (Only if the mean transverse flow velocity of
$\bar v_\perp$=0.5\,$c$ had been established directly after impact,
the matter could have expanded transversely from 4.5~fm to about 9~fm
during the time $\langle\tau\rangle=8.4\, \mbox{fm}/c$.)
The important 
implication is that, although the reaction zone appears to be 
expanding boost-invariantly at freeze-out, it cannot have expanded 
so rapidly throughout its history. {\em The fireball must have 
undergone longitudinal acceleration.} Again, this should not surprise 
anybody: the measured transverse flow must have been created by 
transverse pressure gradients, and since pressure is locally 
isotropic, it must have also pushed longitudinally. The longitudinal 
velocity gradient measured by $\Rl$ is thus only a snap-shot at 
freeze-out, and it is very likely that some fraction of the 
longitudinal flow, like the transverse one, has developed 
gradually by work done by the pressure. As sketched in 
Fig.~\ref{expan} this automatically leads to a longer real fireball 
lifetime, $\tau_{\rm real}>\tau_0$, between impact and freeze-out.

\section{The phase-space density}
\label{psd}

From the measured single-particle momentum spectrum and two-particle
correlation function one can infer the phase-space density, spatially 
averaged over the homogeneity volume, of the particles {\em immediately 
after freeze-out} \cite{Ber94}. Hence, this measurement is not only 
sensitive to the absolute multiplicity of the particles, but gives 
also hints about the possible appearence of ``overpopulation'' in some 
parts of phase-space. Such an overpopulation might give rise to 
``pion laser'' phenomena etc. \cite{Pratt}. Since we have seen that 
the pionic phase-space seems to be populated more densely than expected 
in chemical equilibrium, an analysis of the phase-space density appears 
to be of interest. 

In this section we first elaborate on the formalism developed in 
\cite{Ber94} and then apply it to the data. A very qualitative study 
\cite{Fer99} of a wide set of different measurements did not indicate 
a large overpopulation of the pionic phase-space, but showed 
signatures of the transverse expansion in form of a ``flattening'' 
of the $m_\perp$-dependence of the (position-) averaged phase-space 
density. Here we want to perform, for the specific set of data 
analyzed in this paper, a more quantitative study of both these 
phenomena. Unfortunately, the data quality does not yet allow 
for high precision investigations; we have to keep this in mind and 
will stay on a rather superficial level.

\subsection{Formalism}
\label{psd-form}

Combining the expression for the two-particle correlation function
\beq
  C(p_1,p_2) = 1 + \frac{\left | \int d^4x\, e^{iq\cdot x}\,
  S(x,K)\right|^2}{\int d^4x\, S(x,p_1)\, \int d^4y\, S(y,p_2)}\, ,
  \label{psd1}
\eeq
($q = p_1 - p_2$, $K=(p_1 + p_2)/2$) with that of the one-particle
spectrum (\ref{rel1}), we obtain
\beq
  P_1(p_1)\, P_1(p_2)\, [C(p_1,p_2) - 1] = \left | \int d^4x\, 
  e^{iq\cdot x}\, S(x,K) \right |^2 .
  \label{psd2}
\eeq
Let us introduce the {\em time integrated emission function} 
\beq
  \label{psd3}
  \Sigma(\bbox{x}, K) = \int_{-\infty}^{\infty} dt\, 
  S(t,\bbox{x}+\bbox{v} t,K) \, ,
\eeq
where $\bbox{v} = \bbox{K}/K^0$.
This quantity allows us to rewrite the integral of (\ref{psd2})
over on-shell momenta $q$ satisfying $q\cdot K=0$,
\bea
\label{psd4}
  &&\int d^4q\, \delta(q\cdot K) \left [ P_1(K+{\textstyle{q\over 2}})\, 
  P_1(K-{\textstyle{q\over 2}})\, 
  (C(q,K) - 1) \right ] 
\nonumber \\
  && \qquad 
  \approx \frac{(2\pi)^2}{E_K} \int d^3x \, \Sigma^2(\bbox{x}, K)\, .
\eea
Here, the on-shell approximation (which is valid for $q^2 \ll 4E_K^2$)
has allowed the replacement $K^0 \to E_K= \sqrt{m^2 + \bbox{K}^2}$. We
also find
\beq
  \label{psd5}
  P_1(p) = \int d^3x\, \Sigma(\bbox{x},p) \, .
\eeq

$\Sigma(\bbox{x},p)$ is not the phase-space density. In the
following we establish how it is connected to the latter.
The phase-space density $f(t,\bbox{x},\bbox{p})$ is obtained by
summing over all particles of a given momentum emitted up to 
time $t$ along a given trajectory:
\beq
  \label{psd6}
  f(t,\bbox{x},\bbox{p}) = \frac{(2\pi)^3}{E_p}\, \int_{-\infty}^t dt'\,
  S(t',\bbox{x}+\bbox{v}(t' - t),\bbox{p})\, .
\eeq
The factor in front of the integral assures the correct normalization of
$f$ to the number of particles for $t>t_f$, where $t_f$ is the last
instant of the freeze-out process.

One easily can show \cite{diss} that
\beq
  \label{psd7}
  \frac{E_p^n}{(2\pi)^{3n}} \int d^3x\, 
  f^n(t>t_f,\bbox{x}, \bbox{p}) = \int d^3x\, 
  \Sigma^n(\bbox{x},\bbox{p})\, .
\eeq
From (\ref{psd4}) and (\ref{psd5}) then follows 
\bea
  \label{psd8}
  &&\int d^4q\, \delta(q\cdot K)\, \left [ P_1^2(K)( C(q,K) - 1) \right ]
  \nonumber \\
  && \qquad  \approx  \frac{E_K}{(2\pi)^3} \int d^3x \, 
  f^2(t>t_f,\bbox{x},\bbox{K}) \, , \\
  \label{psd9}
  &&P_1(\bbox{K}) = \frac{E_K}{(2\pi)^3} \int d^3x \, 
  f(t>t_f,\bbox{x},\bbox{K}) \, .
\eea
In (\ref{psd8}) we have performed the smoothness approximation
$P_1(K+{q\over 2})\approx P_1(K-{q\over 2})\approx P_1(K)$. Dividing 
these two equations (and changing the notation $K{\to}p$) we obtain
\bea
\label{psd10}
\langle f \rangle (\bbox{p}) & = & \frac{\int d^3x\,  
        f^2(t>t_f, \bbox{x}, \bbox{p})}{\int d^3x \, 
        f(t>t_f, \bbox{x}, \bbox{p})} \, , \\
\label{psd11}   
& \approx & P_1(p) \int d^4q\, \delta(q \cdot p)\, [C(q,p)-1] \, .
\eea
This allows to determine {\em the phase-space density of free-streaming 
particles averaged over positions at constant global time}, since all 
quantities on the r.h.s.\ can be measured. Due to Liouville's theorem,
the phase-space density of free-streaming particles does not change,
and hence (\ref{psd11}) gives the phase-space density averaged along the
freeze-out hypersurface.

Indeed, for a hypersurface $\sigma_f$ on which the freeze-out process 
is just completed and a global time coordinate is $t_f(\bbox{x})$
one can show \cite{diss} that
\bea
   &&E_p\,\int_{\sigma_t} d^3x\, f^n(t>t_f,\bbox{x},\bbox{p}) 
 \nonumber \\
   &&\qquad\qquad =
   \int_{\sigma_f} p^\mu d^3\sigma_\mu(x)\, 
         f^n(t_f(\bbox{x}),\bbox{x}, \bbox{p})\, ,
 \label{psd12}
\eea
where $\sigma_t$ is the hypersurface given by $t$\,=\,const.\,$>t_f$ 
and $d^3\sigma_\mu$ is the infinitesimal normal vector to $\sigma_f$. 
The factor $p\cdot d^3\sigma$ is known from the formalism of 
Cooper and Frye \cite{CF74} and stands for the flux of the particles 
across $\sigma_f$. This relation allows us to rewrite (\ref{psd10}) as
\beq
  \label{psd13}
  \la f \ra(\bbox{p}) = 
  \frac{\int_{\sigma_f} p\cdot d^3\sigma(x)\, 
             f^2(t_f(\bbox{x}), \bbox{x}, \bbox{p})}
       {\int_{\sigma_f} p\cdot d^3\sigma(x)\, 
             f(t_f(\bbox{x}), \bbox{x}, \bbox{p})}\,.
\eeq
This is the {\em phase-space density averaged over the hypersurface
along which the freeze-out is just completed}. Eq.~(\ref{psd13}) 
establishes what can be learnt about the phase-space density in a 
model-independent way. A back-extrapolation across the freeze-out 
boundary is only possible if additional assumptions about the
mechanism of particle production are made.

If the correlation function is parametrized as in (\ref{rel2}), 
the integration over $q$ in (\ref{psd11}) is simple and leads to 
\bea
  \label{psd99}
  \la f \ra(\bbox{p}) &=& \frac{1}{E_p}\, \frac{1}{p_\perp}\,
  \frac{d^3N}{dp_\perp\, dy\, d\phi}\, 
  \nonumber \\
  &\times& \frac{\pi^{\frac32}}{%
  \Rs(\bbox{p})\sqrt{\Rod(\bbox{p})\Rld(\bbox{p})- (\Rold(\bbox{p}))^2 }}\, .
\eea
However, in this form all pions count towards (\ref{psd99}) irrespective 
of their origin. We want to eliminate the pions from longlived
resonances since they do not contribute to the phase-space density
at freeze-out. (We keep pions from short-lived resonances because these
decay essentially in the same spatial region where the direct pions are 
set free.) The standard procedure for separating off long-lived 
resonances is based on the observation that they lower the 
intercept $\lambda_{\rm dir}$ of the correlator at $q=0$
\cite{CL96a,WH97res}. Assuming that there is no coherent contribution 
to pion emission which would lower the intercept even further \cite{Pra86},
we obtain the phase-space density of ``direct'' pions by multiplying
the r.h.s. of (\ref{psd99}) with $\sqrt{\lambda_{\rm dir}}$ 
\cite{E877,Fer99}. We further simplify (\ref{psd99}) by introducing
the following parametrization for the single-particle spectrum:
\beq
  \label{psd14}
  \frac{1}{p_\perp} \, \frac{d^3N}{dp_\perp\, dy\, d\phi}  = 
  \frac{1}{2\pi}\, \frac{dN}{dy}\, \frac{1}{T^2_{\rm inv}(y)}\,
  \exp\left ( - \frac{p_\perp}{T_{\rm inv}} \right )\, .
\eeq

The final formula used in the following  data analysis then reads 
\bea
  \la f \ra(\bbox{p}) &=& \frac{\sqrt{\pi}}{2}\, 
  \frac{\sqrt{\lambda_{\rm dir}}}{E_p\, T_{\rm inv}^2(y)}\,
  \exp\left ( - \frac{p_\perp}{T_{\rm inv}} \right )\, 
  \frac{dN}{dy}
  \nonumber \\
  &\times& \frac{1}{\Rs(\bbox{p})\sqrt{\Rod(\bbox{p})\Rld(\bbox{p})
  - (\Rold(\bbox{p}))^2 }} \, .
  \label{psd15}
\eea

\subsection{Application}
\label{psd-app}
\subsubsection{Data choice}
\label{psd-app-dat}

The rapidity distribution $dN/dy$ is estimated from results given in
\cite{NA49stop,Gunther} to be $139\pm 22$. An estimate for
 the 
inverse slope is obtained from the measured $\la p_\perp \ra$
\cite{Jones} as
$T_{\rm inv} = 185 \pm 10\, \MeV$. The correlation radii are taken 
from Table~\ref{danal-bpdata}. Since for the investigated rapidity bin 
no intercept parameter was given for the MTPC analysis~\cite{Stefan}
we estimated it by averaging only the VTPC data for negative and 
positive hadrons \cite{Harry}. As seen in Table~\ref{psd-t1} they 
are fairly $K_\perp$-independent; moreover they agree well with the 
intercept parameters extracted in \cite{Stefan} for $4{<}Y{<}5$.

\subsubsection{Phase-space densities from statistical distributions
and realistic emission functions}
\label{psd-app-dist}

\begin{figure}
 \begin{center}
 \epsfxsize=8.6cm
 \centerline{\epsfbox{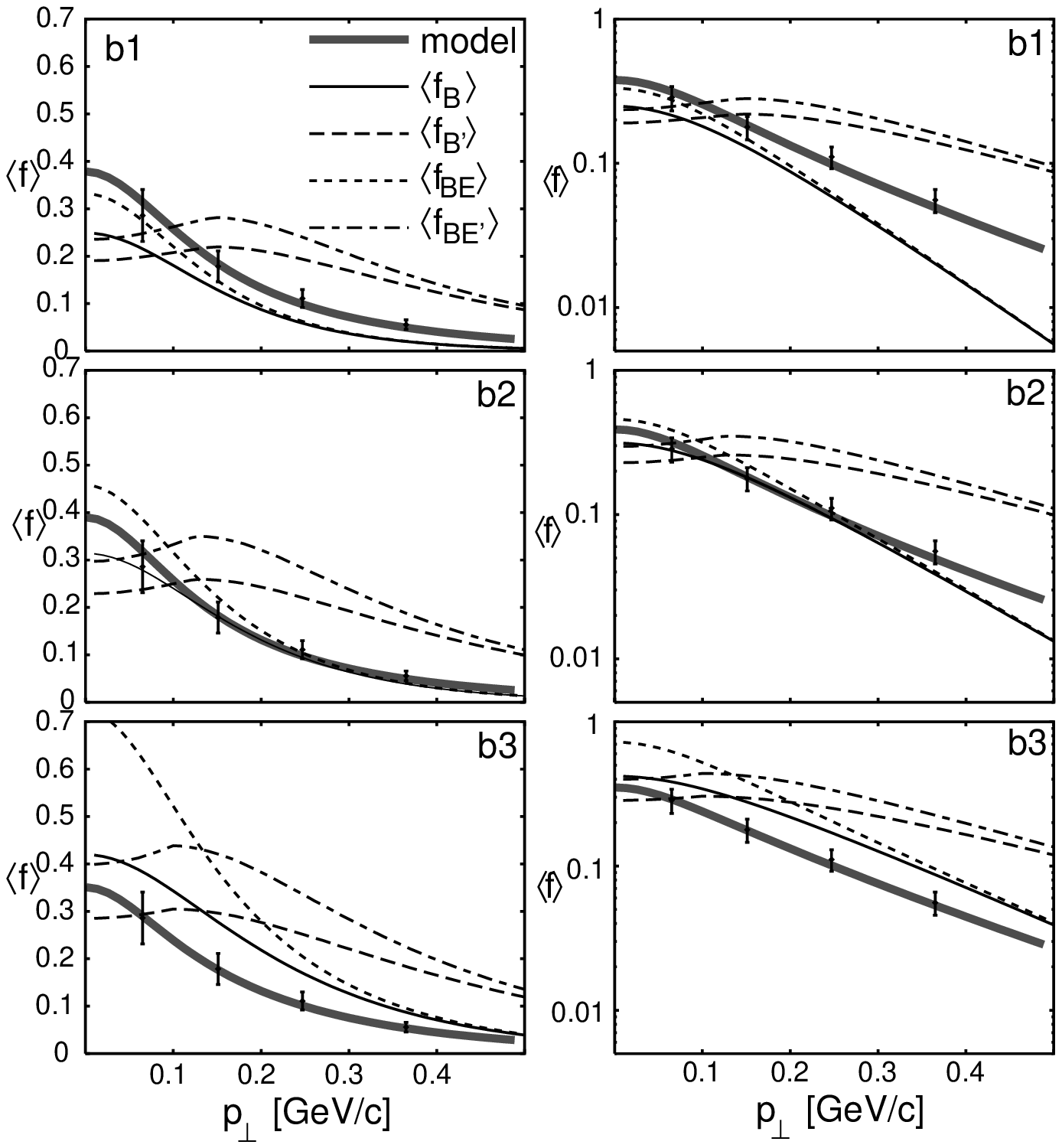}}
\caption{Average phase-space density $\la f \ra(\bbox{p})$ in the rapidity 
        bin $3.9<Y<4.4$ as a function of $p_\perp$. The data points are 
        determined from NA49 data (Sec.~\ref{psd-app-dat}). Black 
        lines denote expectations from the statistical treatment in 
        Sec.~\ref{psd-app-dist}: $\la f_B \ra$ (eq.\ (\ref{psd17}), solid);
        $\la f_{B'} \ra $ (eq.\ (\ref{psd19}), long-dashed); 
        $\la f_{BE} \ra$ (eq.\ (\ref{psd18}), short-dashed); and
        $\la f_{BE'} \ra$ (eq.\ (\ref{psd20}), dash-dotted). Results of
        our model (see text) are shown by thick solid gray lines. The symbols
        in the upper right corners indicate the used set of model 
        parameters; the corresponding pionic chemical potentials are 
        listed in Table~\ref{psd-t2}. The right column gives a logarithmic
        representation of the left column.
\label{psd-f2}}
\end{center}
\vspace*{-0.6cm}
\end{figure}

We first study the simple question to what extent the data support
the simple assumption that the observed phase-space density follows 
a purely quantum-statistical distribution. Rough agreement of the data 
with a Bose-Einstein distribution was observed in \cite{Fer99}. Since 
our model reproduces both the measured one- and two-particle spectra, 
it allows us to refine these observations. To this aim, we compare 
in a first step in Fig.~\ref{psd-f2} the data with simple expectations 
from statistical distributions.

Since in LCMS the energy coincides with the transverse mass
$E=m_\perp$, the phase-space occupancy following from a Boltzmann
distribution is simply
\beq
\label{psd17}
\la f_B \ra (\bbox{p}) = \exp \left (-\frac{m_\perp}{T} \right )\, ,
\eeq
while the expectation based on the Bose-Einstein distribution reads
\beq
\label{psd18}
\la f_{BE} \ra (\bbox{p}) = \frac{1}{e^{\frac{m_\perp}{T}} -1} \, .
\eeq
The temperatures for the corresponding curves in Fig.~\ref{psd-f2} are 
taken from the parameter sets indicated in the upper right corner. One 
sees that unless the temperature reaches $160\, \MeV$ (which is
disfavoured by the fits) the slope of the calculated distributions 
is steeper than that of the data.

The results of Sec.~\ref{fit} identify as the main source of this 
discrepancy the strong transverse flow at freeze-out. That 
transverse flow has a strong effect on $\la f \ra (\bbox{p})$
can be seen already from a minimal modification of (\ref{psd17}), 
(\ref{psd18}), in which the statistical distributions are ``boosted'' 
to the rest frame of the point of maximum emissivity for a given 
flow field, $u_{\rm max}$:
\bea
\label{psd19}
\la f_{B'} \ra(\bbox{p})& =& \exp \left ( - \frac{p\cdot u_{\rm max}}{T}
\right ) \, ,\\
\label{psd20}
\la f_{BE'} \ra (\bbox{p}) &=& \frac{1}{\exp \left (
\frac{p\cdot u_{\rm max}}{T} \right ) -1} \, .
\eea

We note that the shape of the expectations $\la f_{B'} \ra(\bbox{p})$ 
and $\la f_{BE'} \ra (\bbox{p})$ (e.g.\ ``the bump'') shown in 
Fig.~\ref{psd-f2} depend on the determination of $u_{\rm max}$ which 
is model-dependent. For instance, the bump vanishes if a Gaussian 
transverse density profile is taken, while the slope shows a 
qualitatively similar behaviour \cite{diss}. In any case, as seen 
in Fig.~\ref{psd-f2}, the statistical distributions (\ref{psd19}) 
and (\ref{psd20}) overpredict the phase-space density significantly 
for large transverse momenta. A more refined model is needed to 
account for the data.

This was done, using the model described in Sec.~\ref{model} to 
calculate the HBT radius parameters and the one-particle spectra 
which enters the numerator of Eq.~(\ref{psd99}). As mentioned above, 
the data are multiplied with $\sqrt{\lambda_{\rm dir}(\bbox{p})}$ and 
measure the phase-space occupancy of {\em direct pions} only. 
Furthermore, without chemical potentials, our model leads to smaller 
multiplicities than measured. This allows us to determine the chemical 
potential needed in order to reproduce the observed phase-space density.
If a
non-zero chemical potential is considered, the difference 
between
the Bose-Einstein distribution and its Boltzmann approximation
increases. Therefore, for the calculation of the single-particle
spectrum the Bose-Einstein distribution was implemented in an
analogous way as described in Appendix \ref{mult-be}. 

As seen in Fig.~\ref{psd-f2}, the transverse momentum
dependence of the phase-space density is well described with our model
whose geometrical and dynamical input was extracted from a space-time
analysis of hadronic spectra. The chemical potentials needed to account 
for the absolute particle yields are listed for the three different
parameter sets in Table~\ref{psd-t2}. In contrast to the box-shaped 
density profile, the multiplicities obtained from a Gaussian one 
require chemical potentials above 135 MeV. This may be taken as yet 
another hint that a box-shaped density profile is favoured by the data. 
Also, at such high values of $\mu$ the effects due to the finite fireball 
volume may become relevant and the treatment gets more complicated 
\cite{claus}.

\subsubsection{Discussion}
\label{psd-app-diss}

The determination of the pionic chemical potential as a measure
of the ``overpopulation'' of phase-space is of interest for the 
discussion of generic quantum-mechanical effects (e.g. stimulated 
emission \cite{PrattQGP2}), for the study of the fireball's
chemistry \cite{Beb92}, as well as for certain signals of 
in-medium changes of hadron masses \cite{Koch,PH98,VCK98}. Here 
we discuss the implications of Fig.~\ref{psd-f2} in this context.

As noted above, the model curves in Fig.~\ref{psd-f2}
reproduce the transverse momentum dependence better than 
simple statistical distributions. However, they account for 
the absolute particle yields only with the help of a relatively
large chemical potential while the statistical distributions
rather overestimate the absolute yields without invoking a chemical 
potential. The latter effect can be understood by recalling that the
distributions (\ref{psd17}) and (\ref{psd18}) were evaluated with
energy distributions at the point of maximum emissivity. They thus 
overestimate the data by construction. To understand the rather
large chemical potentials required for our model, we recall that 
in the present study the value of this chemical potential is subject
to other model-dependent features:
\begin{enumerate}
\item
The $\eta$-dependence of the emission function (\ref{mod1})
introduces the factor $\exp[-(\eta{-}\eta_0)^2/(2(\Delta\eta)^2)]$.
For the rapdity bin $3.9{<}y{<}4.4$, this effectively amounts to a
negative chemical potential of $-0.46\,T$. For the temperatures 
from sets {\sf b1} and {\sf b2} this gives $\approx -50$\,MeV which 
must be compensated by the ``usual'' chemical potential. Note that 
this compensation would not be required if a box-shaped density 
profile were also used for the $\eta$-distribution of the source; such 
a distribution seems also to be consistent with the recent study of the 
phase-space density in Pb+Pb collisions \cite{Fer99}.
\item
The fitted chemical potential in Table~\ref{psd-t2} also accounts 
implicitly for the inclusion of pions from short-lived resonances. For 
high temperatures ($\sim 150\, \MeV$) this is an important effect, 
since then the contribution by these resonances is comparable with 
direct pion production. The exact fraction of particles from resonance 
decays depends on the chemical potentials for the resonances and requires
a more detailed picture of the fireball chemistry. It decreases with
decreasing temperature \cite{SKH91}. For a simple estimate one may
count the number of pions from short-lived resonances as calculated
from our model. The pion chemical potential needed to compensate the
lack of these resonance contributions in our emission function by 
an additional amount of direct pions is $\mu \approx 20\, \MeV$ for 
the set {\sf b1} and $\mu \approx 34\, \MeV$ for the set {\sf b2}. 
The higher value of $\mu$ in the latter case reflects the larger 
resonance fraction at the higher temperature.
\end{enumerate}

Subtracting these two effects from the values given in Table~\ref{psd-t2}
we are led to a ``real pion chemical potential'' $\mu_\pi$ which is 
significantly smaller: we find $\mu_\pi \approx 60\, \MeV$ for 
$T=100$\,MeV (set {\sf b1}) and $\mu_\pi \approx 25\, \MeV$ for 
$T=120$\,MeV (set {\sf b2}). These parameters still indicate 
a slight overpopulation of the pion phase-space at freeze-out. Clearly, 
our discussion could be refined by a proper study of the rapidity 
dependence of the particle spectra and yields and by a more 
quantitative inclusion of contributions from short-lived resonances.
This, however, lies outside the scope of the present work.

\section{Conclusions}
\label{summary}

In the context of simple analytical parametrizations of the freeze-out 
emission function, we have presented a space-time analysis of hadronic 
one- and two-particle spectra measured for slightly forward rapidity
by the NA49 Collaboration at the CERN SPS. Our work quotes for the first time
full $\chi^2$ confidence levels for the temperature and the transverse
flow. It allows for a quantitative discussion of possible
model-dependencies via the comparison of different transverse density
profiles. Thus, our results corroborate quantitatively the picture of
a collision system with strong transverse collective flow which drives
an expansion to twice the initial transverse size before
freeze-out. During the evolution to this freeze-out stage the system
seems to cool down significantly. 

Most importantly,
this general picture does not depend on details of the model emission
function with which the analysis is done. The $\chi^2$ values of the
fit are, however, sensitive to details of the model. With the used
transverse flow profile they 
favour a box-shaped over a Gaussian transverse density profile.
This gives support to a similar conclusion reached recently in the 
study of deuteron coalescence models, and it indicates a possible 
different origin for negative $\la \tilde{x}^2 - \tilde{y}^2\ra$
contributions previously attributed to opacity effects.

The strong $M_\perp$-dependence of the HBT side and out radius 
parameters requires a large transverse flow and thus favours a low
kinetic freeze-out temperature in the range $80\, \MeV < T < 110 \, \MeV$. 
Other earlier analyses which had obtained higher freeze-out temperatures
therefore require some discussion:
\begin{itemize}
\item
In \cite{kampfer} a temperature $T = 120\, \MeV$ combined with 
$\bar v_\perp = 0.43$ was found from a simultaneous fit to 
single-particle $m_\perp$-spectra of different species. This value
does not contradict our findings since it lies only slightly below 
the 90\% confidence level of our fit to correlation radii.
(The combination of transverse velocity and density profile used 
in \cite{kampfer} differs from ours. This can lead to slightly 
different fit results). While a simultaneous analysis of only the 
single-particle spectra of many different particle species thus 
also permits the separation of $T$ and $\bar v_\perp$, it must make 
the additional assumption that all these hadron species decouple 
simultaneously. That one obtains in this way similar results as
by combining spectra and correlations of a single particle species 
(pions) suggests that this assumption is in fact reasonably well 
justified.

\item
By studying the curvature of the very accurately measured $\pi^0$ 
spectrum in $m_\perp$ up to 4~GeV/$c$ the WA98 collaboration extracted
an apparently highly accurate value of the freeze-out temperature of 
about $185\, \MeV$ \cite{peitzmann}. This analysis was criticized 
\cite{UrsQM99} on the basis that the curvature in the high 
$p_\perp$ region results from the Cronin effect and that an agreement 
with a thermal model is likely to be an artefact and is subject to 
severe systematic uncertainties in the model parametrization.
\item
The NA49 collaboration has analyzed their data on the basis of 
simplified analytical expressions derived in \cite{CNH95},
such as e.g. Eq.~(\ref{dan4}). In this way they arrived at
$T \approx 120\, \MeV$ and $\bar v_\perp \approx 0.55$
\cite{App98}. The discrepancy with our fit parameters can 
be traced back to the limited validity of these analytical 
expressions, especially for the description of the $h^-$ spectrum.
If one compares their corresponding $\chi^2$ valley with ours
(Figs.~\ref{danal-f3} and \ref{danal-f4}) one sees that it bends 
over at low $T$ and large $\bar v_\perp$ whereas ours doesn't. This
is the main difference and accounts for their intersection to occur 
at somewhat larger values of $T$ and $\bar v_\perp$.
\item
An analysis in the spirit of ours was reported in \cite{Ster}. Again,
higher temperatures (134--145~MeV depending on the analyzed set of data)
were quoted. In this case we suspect that the high temperature 
values are driven by requiring a fit to the {\em normalized} 
single-particle spectra {\em with vanishing pion chemical potential}. 
The high temperature is thus a consequence of the observed multiplicity. 
Moreover, the correlation radii were calculated from analytical 
approximations \cite{CL96a} which again introduce additional 
uncertainties.
\end{itemize}

Our analysis of the observed phase-space density (\ref{psd15}) has
shown conclusively that the main deviation of the experimental data
from a purely quantum-statistical phase-space distribution is due
to the collective dynamical expansion of the system. Also, we have 
inferred from our analysis a pion chemical potential $\mu_\pi
 \approx 60\, \MeV$ at $T= 100\, \MeV$. This indicates an 
overpopulation of the pion phase-space by about a factor 2 at 
thermal freeze-out, consistent \cite{Beb92} with analyses of the
particle ratios indicating a much earlier decoupling of the hadron 
yields (chemical freeze-out), namely at $T_{\rm chem} \approx 170$ 
MeV in Pb+Pb collisions at the SPS. Nevertheless, this pion excess
is too small to necessitate the study of stimulated pion emission 
\cite{PrattQGP2} or higher order correlations \cite{UrsCorr}. 


\acknowledgements
We thank P.~Seyboth and R.~Stock for the permission to use 
unpublished NA49 data and H.~Appelsh\"auser, R.~Ganz and 
S.~Sch\"onfelder for providing us with these data. We are
indebted to H.~Appelsh\"auser, J.G.~Cramer, D.~Ferenc, H.~Heiselberg, 
T.~Hemmick, R.~Matiello, A.~Polleri, J.~Sollfrank, H.~Sorge, P.~Seyboth 
and R.~Stock for valuable discussions. Our work was supported by 
grants from DAAD, BMBF, DFG, GSI, and the US Department of Energy.

\appendix


\section{Distribution of the pair momentum}
\label{pmom}

In this Appendix we derive distributions of the pair rapidity $Y$ 
and of the pair transverse momentum $K_\perp$.

The observed $h^-$ single-particle rapidity spectrum can be very
accurately approximated by a Gaussian distribution of width 
$\Delta y=1.4$ \cite{NA49stop,Gunther}:
\beq
  \label{pd1}
  \varrho_1^y(y) = \frac{1}{\sqrt{2 \pi (\Delta y)^2}}
  \, \exp\left (-\frac{(y - y_0)^2}{2 (\Delta y)^2} \right ) \, . 
\eeq
Here $y_0$ stands for midrapidity, and the distribution is normalized
to unity (rather than to the total number of particles). The distribution 
of the pair rapidity 
\beq
  \label{pd2}
  Y = \frac{y_1 + y_2}{2}
\eeq
is then evaluated as
\bea
 \label{pd4}
   \varrho_2^Y(Y) &=& \int dy_1\, dy_2\, 
   \delta\! \left (Y - \frac{y_1+y_2}{2}\right )\, 
   \varrho_1^y(y_1)\, \varrho_1^y(y_2) 
 \nonumber\\
   &=& \frac{1}{\sqrt{\pi\, (\Delta y)^2}} \, 
   \exp\left ( - \frac{(Y - y_0)^2}{(\Delta y)^2} \right ) \, .
\eea

When deriving the analogous distribution of the mean transverse 
momenta $K_\perp$ of the pairs one starts from the corresponding 
single-particle distribution (again normalized to unity) \cite{Gunther}
\beq
  \label{pd5}
  \varrho_1^\perp(p_\perp) = \frac{1}{\cal N}\, \frac{d^2N}{d\phi\, dp_\perp} =
  \frac{1}{2\pi}\, \frac{p_\perp}{T_{\rm inv}^2}\, \exp\left (- 
  \frac{p_\perp}{T_{\rm inv}} \right ) \, ;
\eeq
here $T_{\rm inv}$ is the measured inverse slope, and ${\cal N}$ is a 
normalization constant. The average transverse momentum of a pair with 
transverse momenta $\bbox{p}_{1\perp},\bbox{p}_{2\perp}$ is 
\beq
  \label{pd6}
  K_\perp^2 = \frac14 \left ( p_{1\perp}^2 + p_{2\perp}^2 + 
  2 p_{1\perp} p_{2\perp} \cos(\phi_1 - \phi_2) \right )\, ,
\eeq
where $\phi_1$ and $\phi_2$ are the azimuthal angles of the individual
transverse momenta. Then the desired distribution of $K_\perp$ is
obtained from
\bea
  \varrho_2^\perp(K_\perp) &=& \int_0^\infty dp_{1\perp} \int_0^{2\pi} d\phi_1
  \int_0^\infty dp_{2\perp} \int_0^{2\pi} d\phi_2\, 
  \nonumber \\
  && \times
  \varrho_1^\perp(p_{1\perp})\, \varrho_1^\perp(p_{2\perp})\, 
  \delta\left(K_\perp - \bar{K}_\perp\right)\, , 
  \label{pd7} \\  
  \bar{K}_\perp &=& 
  \frac12 \sqrt{p_{1\perp}^2 + p_{2\perp}^2 + 2p_{1\perp} p_{2\perp}
  \cos(\phi_1 - \phi_2)}\, .
  \label{pd7b}
\eea

We now integrate over the angles with the help of the $\delta$-function.
The integration boundaries are most easily implemented by using as 
integration variables
\begin{mathletters}
\label{pd11}
\bea
\label{pd11a}
\frac12(p_{1\perp} + p_{2\perp}) & = & \zeta \in (K_\perp, \, \infty)\, , \\
\label{pd11b}
\frac12(p_{1\perp} - p_{2\perp}) & = & \xi \in (-K_\perp,\, K_\perp)\, .
\eea
\end{mathletters}
Integrating over $\xi$ and substituting $\zeta$ by a 
dimensionless variable $\alpha = \zeta/K_\perp$ we obtain the
final expression \cite{diss}
\beq
 \label{pd13}
 \varrho_2^\perp(K_\perp) = 2\, \frac{K_\perp^3}{T_{\rm inv}^4}\,
 \int_1^\infty d\alpha \, 
 \exp\left (-\frac{2K_\perp}{T_{\rm inv}}\alpha \right )\,
 \frac{2\alpha^2 - 1 }{\sqrt{\alpha^2 - 1}} \, .
\eeq

\begin{figure}
 \begin{center}
 \epsfxsize=7.1cm
 \centerline{\epsfbox{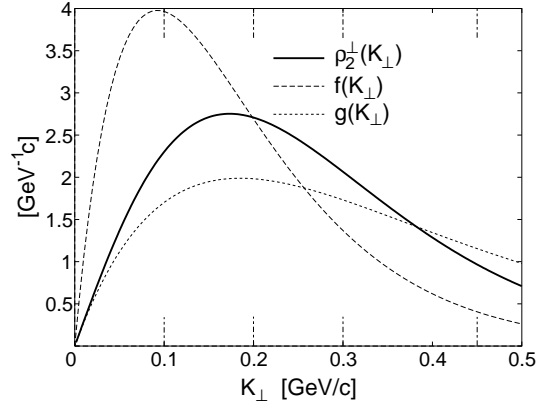}}
 \caption{The distribution $\varrho_2^\perp(K_\perp)$ of average 
    transverse pair momenta calculated from (\ref{pd13}). The dashed 
    line shows the function $f(K_\perp)$ from Eq.~(\ref{pd14}); the
    dotted line stands for $g(K_\perp)$ from (\ref{pd15}). 
    $T_{\rm inv} = 185\, \MeV$. Longer tics indicate the bin positions
    for the data in Sec.~\ref{data}.
 \label{ktdist}}
\end{center}
\vspace*{-0.5cm}
\end{figure}

\begin{figure}
 \begin{center}
 \epsfxsize=7.1cm
 \centerline{\epsfbox{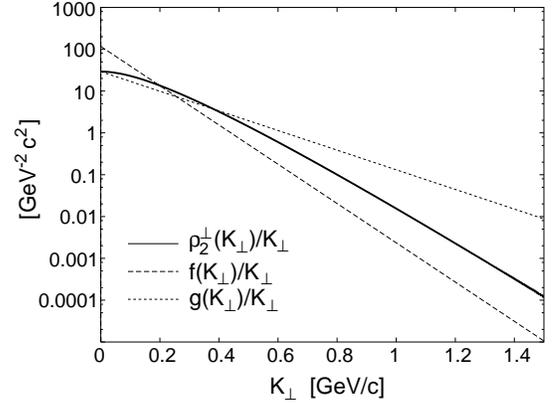}}
 \caption{The same as Fig.~\ref{ktdist}, but all three distributions
        are divided by $K_\perp$ and plotted logarithmically.}
 \label{ktdlog}
\end{center}
\end{figure}

The resulting distribution, for an inverse slope $T_{\rm inv} = 185$\,MeV 
as used in the calculations, is illustrated in Fig.~\ref{ktdist}.
It is compared with the two naive guesses (both normalized to unity)
\begin{mathletters}
\label{pd1415}
\bea
f(K_\perp) & = & \frac{4K_\perp}{T_{\rm inv}^2} \, \exp\left ( - 
\frac{2K_\perp}{T_{\rm inv}} \right ) \, , 
\label{pd14} \\
g(K_\perp) & = & \frac{K_\perp}{T_{\rm inv}^2} \, \exp\left ( - 
\frac{K_\perp}{T_{\rm inv}} \right ) \, .
\label{pd15}
\eea \end{mathletters}
Note that $g(K_\perp)$ is actually the distribution of $p_\perp$ 
integrated over the azimuthal angle (cf.\ (\ref{pd5})); it seems 
to approximate $\varrho_2^\perp(K_\perp)$ better than $f(K_\perp)$. 
However, as seen in Fig.~\ref{ktdlog} this is only the case for 
$K_\perp < 500\, \MeV/c$. For high $K_\perp$ the distribution 
$\varrho_2^\perp(K_\perp)$ asymptotically behaves like $f(K_\perp)$. 
Thus in general, a replacement of $\varrho_2^\perp(K_\perp)$ by 
either of the two distributions (\ref{pd1415}) cannot be re\-com\-mended.

\section{Computation of the negative hadron transverse momentum spectrum}
\label{comps}

The theoretical $p_\perp$-spectrum needed in the fit in 
Section~\ref{sinspec} was calculated with the routine described in
\cite{WH97res}. Here we point out the changes in the original routine
which were made before it was used in this work.

\subsection{Bose-Einstein distribution for direct pions}

The directly produced pions are distributed according to the Bose-Einstein
distribution, unlike in the original code in which the Boltzmann
(high-energy) approximation was employed. In practice this was done 
by expanding the Bose-Einstein distribution into powers of the Boltzmann
distribution and truncating after a sufficient number of terms
(see also Appendix~\ref{mult-be}). The pion chemical potential was set
to zero. 

\subsection{Chemical potentials for resonances}

Baryon number and strangeness conservation in the fireball requires 
the introduction of nonzero baryon and strangeness chemical potentials 
$\mu_B$ and $\mu_S$. Since these chemical potentials turn out to be not
negligible \cite{Sta99,Shur,BGS98,Bra98}, they may affect the
multiplicities of baryonic and/or strange resonances and thus also the
number of pions. The condition of strangeness neutrality of the 
fireball determines $\mu_S$ as a function of $T$ and $\mu_B$.

\begin{figure}
\begin{center}
 \epsfxsize=7.0cm
 \centerline{\epsfbox{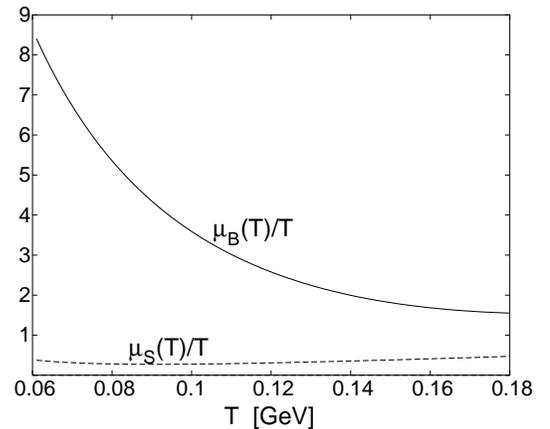}}
\caption{The chemical potentials $\mu_B(T)$ and $\mu_S(T)$ in units 
   of $T$ as given by Eqs.~(\ref{chempot}) and (\ref{stpot}). 
  \label{fuga}}
\end{center}
\vspace*{-0.8cm}
\end{figure}

The yield of particles with strangeness $S$ is then multiplied by a 
factor $\lambda_S = \exp(S\,\mu_S(T)/T)$. Guided by the
dependence 
of $\mu_B$ on the temperature shown in Fig.~3 of Ref.~\cite{Shur} and 
with an eye on the results of \cite{Bra98,Sta99} we introduced a 
polynomial parametrization for $\mu_B(T)$: 
\beq
 \label{chempot}
   \mu_B(T) = a\, T^2 + b\, T + c\, ,
 \eeq
with $a = 24.8\, \GeV^{-1}$, $b = -7.94$, and $c = 0.905\, \GeV$. 
Although rough, this parametrization is sufficiently precise 
for our purposes and easy to handle. The strangeness chemical 
potential $\mu_S$ was calculated from
 \beq
 \label{stpot}
   0 = \sum_i S_i\, g_i\, m_i^2\, T\, e^{(B_i\mu_B+S_i\mu_S)/T} 
   \bessk{2}\left(\frac{m_i}{T}\right)
   \, ,
 \eeq
where $T$ is the freeze-out temperature and the sum runs over all 
species. Particle masses, baryon numbers and strangeness are denoted 
by $m_i$, $B_i$, and $S_i$, respectively; $g_i$ is the spin and isospin 
degeneracy. When determining 
$\mu_S(T)$ all resonances up to 2~GeV were taken into account, in order 
to be consistent with \cite{Shur}. The resulting dependences of the 
chemical potentials on the temperature are shown in Fig.~\ref{fuga}.

\subsection{Non-pion contributions and resonance decays}

Since we studied data on $h^-$-spectra rather than identified pions, 
our calculations had to include non-pion negative hadrons. Their 
distribution is assumed to be given by the same emission function 
with appropriately modified masses and multiplied by the corresponding
spin-isospin degeneracy factor. We used the Boltzmann approximation, 
which is justified due to the large masses of these particles,
with the appropriate chemical potentials.

Some kaons and antiprotons are also produced by resonance decays. We
included the same set of resonances as in the calculation of the pion
spectrum (see Table~I of \cite{WH97res}). The decays contributing to 
the $K^-$ spectrum are listed in Table~\ref{comps-K}, those leading
to antiprotons in Table~\ref{comps-p}. 

As in \cite{WH97res}, contributions to the pion spectra from 
$\Sigma$ and $\Lambda$ decays were treated as decays of a single 
resonance $Y$ with an average mass of $1.15\, \GeV$. Antiproton 
production via decay chains of the type $\bar\Sigma^* \to \bar Y
 + \ldots \to \bar p + \ldots$ was effectively included by enhanced 
branching ratios for the $\bar Y$ decay channels. (We have also 
replaced by $J=3/2$ the erroneous value $J=1/2$ for the spin 
degeneracy of $\Sigma^*$ listed in \cite{WH97res}.) As argued in 
\cite{WH97res}, the above approximations work well because they 
only concern small relative contributions to the full result.
\section{Total pion multiplicity}
\label{mult}

\subsection{Boltzmann distribution}
\label{mult-boltz}

In this Appendix we derive formulae for the total number of produced 
pions predicted by the model (\ref{mod1}), both 
for the Gaussian (Eq.~(\ref{mod6})) and the box-shaped 
(Eq.~(\ref{mod7})) transverse 
density profiles. They are obtained by integrating the emission 
function (\ref{mod1}) over positions and momenta:
\bea
  &&\la N \ra  = \frac{1}{(2\pi)^3}\int p_\perp\, dp_\perp\, dy\, d\phi\, 
  d\varphi\, r\, dr\, d\eta\, \frac{\tau\,d\tau}{\sqrt{2\pi (\Delta\tau)^2}}
  \, m_\perp 
  \nonumber \\
  && \quad \times \cosh(y-\eta)\,
  \exp \left(-\frac{(\tau-\tau_0)^2}{2 (\Delta\tau)^2} \right )\, 
  \exp\left (-\frac{(\eta-\eta_0)^2}{2 (\Delta\eta)^2} \right )
  \nonumber \\
  &&\quad \times  
  \exp\left( -\frac{m_\perp\, \cosh(y-\eta) \,
      \cosh(\eta_f r/r_{\rm  rms})}{T} \right) 
  \nonumber \\  
  &&\quad \times G(r)\, \exp\left(\frac{p_\perp\, 
      \sinh(\eta_f r/r_{\rm  rms})\, \cos(\phi -\varphi)}{T}  
  \right)\,  .
  \label{mult1}  
\eea
The resulting expression takes the form \cite{diss}
\beq
  \label{mult-d1}
  \la N \ra = \frac{1}{2\pi^2}\, m^2 \, T\, \bessk{2} \left ( \frac{m}{T} 
  \right )\, V_{\rm inv}\, ,
\eeq
where $V_{\rm inv}$ is the invariant volume of the fireball which
takes into account the Lorentz-contraction of the moving volume 
elements and the Cooper-Frye-like flux through the freeze-out
hypersurface \cite{CF74}. For a sharp freeze-out hypersurface 
$\sigma_f$ at $\tau= \mbox{const.}$, for example, this leads to
\beq
  \label{mult-d2}
  V_{\rm inv} = \int_{\sigma_f} dV_{\rm inv} = 
  \int_{\sigma_f} u \cdot d^3\sigma\, ,
\eeq
with
\beq
  \label{mult-d3}
  d^3\sigma_\mu = (\cosh\eta,\, 0,\, 0,\, \sinh\eta)\, \tau\, d\eta\, 
  r\, dr\, d\varphi
\eeq
and (using (\ref{mod8}))
\beq
  \label{mult-d4}
  dV_{\rm inv} = u \cdot d^3\sigma =
  \cosh\eta_t \, \tau\, d\eta\, r\, dr\, d\varphi\, .
\eeq
The model (\ref{mod1}) can be thought of as a Gaussian superposition 
of $\tau= \mbox{const.}$ hypersurfaces. Moreover, the volume is not 
populated equally densely and the corresponding distributions must 
also be taken into account:
\bea
 \nonumber
   V_{\rm inv} &=& 2\pi\int r\, dr\, \cosh\eta_t(r) \, G(r)\,
 \nonumber \\ 
    && \times \int \frac{\tau\, d\tau}{\sqrt{2\pi (\Delta\tau)^2}}\, 
    \exp\left(-\frac{(\tau - \tau_0)^2}{2 (\Delta\tau)^2} \right)   
 \nonumber \\ 
    && \times \int d\eta\,
    \exp\left(-\frac{(\eta - \eta_0)^2}{2 (\Delta\eta)^2} \right)\, .
 \label{mult-d5}
\eea
The last two integrals are trivial, but different transverse density 
and flow profiles lead to different results for the first integral.
The combination of (\ref{mod9}) with (\ref{mod6}) yields
\bea
  V_{\rm inv}^{\rm Gauss} &=& {\textstyle{1\over 2}}(2\pi)^{\frac32}\, 
  r_{\rm rms}^2\, \tau_0\, \Delta \eta\,  
  \nonumber \\
  &&\times \left[1 + {\eta_f\over 2}\, e^{\eta_f^2/4}\, 
                 \int_{-\eta_f/2}^{\eta_f/2} e^{-x^2} \, dx \right ]\, ,
  \label{gauss-vol}
\eea
while the box-shaped transverse density profile gives
\bea
  &&V_{\rm inv}^{\rm box} = (2\pi)^{\frac32}\, r_{\rm rms}^2\,
  \tau_0\, \Delta \eta\,
  \nonumber \\
  &&\quad \times \left [ 
  \frac{\sqrt{2}\sinh(\sqrt{2}\eta_f)}{\eta_f} - 
  \frac{\cosh(\sqrt{2}\eta_f) - 1}{\eta_f^2} \right ]\, .
  \label{box-vol}
\eea
Here $r_{\rm rms} = R_G \sqrt{2}$ for the Gaussian and $r_{\rm rms}
 = R_B/\sqrt{2}$ for the box profile, respectively. For $\eta_f=0$ 
(no transverse flow) the box profile thus gives twice as many pions 
as the Gaussian one (at the same value of $r_{\rm rms}$, i.e.
for the same interferometric signal). This is a volume effect:
at the same $r_{\rm rms}$ the invariant volume is twice as large for 
a box-shaped distribution than for a Gaussian one. 

Transverse flow increases the invariant volume by the factor given 
in the square brackets of (\ref{gauss-vol},\ref{box-vol}). The 
phy\-sical picture is that the transversely moving fluid cells are 
Lorentz contracted and thus more of them are needed to ``fill'' 
the transverse profile of a given size. This Lorentz contraction 
is reflected by the factor $\cosh\eta_t$ (i.e. the $\gamma$-factor 
corresponding to the transverse motion) in (\ref{mult-d4}).
 
\subsection{Bose-Einstein distribution}
\label{mult-be}

Since the pion mass is comparable to the freeze-out temperature,
the use of the Boltzmann approximation for pions is questionable,
in particular if they develop a positive chemical potential. In 
this subsection we show how the formulae for the total particle 
yield are modified if the Bose-Einstein distribution is taken into 
account. In this case the mean pion multiplicity is obtained from
\bea
  && \la N \ra = \int \frac{d^3p}{E} 
  \int \frac{d\tau}{\sqrt{2\pi (\Delta\tau)^2}}\, \exp\left ( 
  -\frac{(\tau - \tau_0)^2}{2 (\Delta\tau)^2} \right ) 
\label{mbe1} 
\\
  && \qquad \times \int_{\sigma_f} p \cdot d^3\sigma \left [ 
  \exp\left( \frac{p\cdot u(x) - \mu_0}{T} - \frac{\mu(x)}{T} \right )
  -1 \right]^{-1} .
\nonumber 
\eea
The last integration gives the momentum spectrum emitted on a
$\tau = \mbox{const.}$ freeze-out hypersurface \cite{CF74}. In the 
second integration one sums up contributions from such hypersurfaces 
with Gaussian distributed $\tau$ values. The ``usual'' chemical 
potential is denoted by $\mu_0$ while the density distribution
of the source is implemented via the ``$x$-dependent part of the 
chemical potential'' \cite{Bol93,Sco98} denoted by $\mu(x)$. The 
model with a Gaussian transverse density profile is thus characterized
by
\beq
\label{mbe2}
\frac{\mu(x)}{T} = - \frac{(\eta - \eta_0)^2}{2 (\Delta\eta)^2} -
\frac{r^2}{2\, R_G^2}\, ,
\eeq
while the box-shaped source is implemented via
\beq
\label{mbe3}
\frac{\mu(x)}{T} = - \frac{(\eta - \eta_0)^2}{2 (\Delta\eta)^2} - B(r)\, ,
\eeq
where 
\beq
\label{mbe4}
B(r) = \left \{ \begin{array}{ll}
                0 & \mbox{if\ }\, r \le R_B, \\
                \infty & \mbox{otherwise.} 
                \end{array} \right. 
\eeq
For practical evaluation the Bose-Einstein term under the last integral 
in (\ref{mbe1}) is expanded into a geometric series. When exchanging 
the order of summation and integrations one arrives at
\bea 
  \nonumber
  &&\la N \ra = \sum_{n=1}^\infty \int \frac{d^3p}{E} \int
  \frac{d\tau}{\sqrt{2\pi\, \Delta\tau^2}}\, 
  \exp\left ( 
  -\frac{(\tau - \tau_0)^2}{2 (\Delta\tau)^2} \right ) \\
  &&\quad \times \int_{\sigma_f} p\cdot d^3\sigma 
  \exp\left(- \frac{p\cdot u(x) - \mu_0}{(T/n)} + 
  \frac{\mu(x)}{(T/n)} \right ) \, .
  \label{mbe5}
\eea
The Boltzmann approximation used in the previous subsection is recovered
from the first term $n=1$. Higher terms $n > 1$ can be regarded as 
Boltzmann contributions with lower effective temperatures $\frac{T}{n}$.
In practice, for sufficiently small chemical potentials, this allows to 
truncate the expansion at a certain value $n_{\rm trunc}$ since the 
size of the contributions decreases exponentially with $n$. For a 
Gaussian density distribution one finds explicitly
\bea
  \la N \ra &= &\sqrt{\frac{2}{\pi}}\, \tau_0\, r^2_{\rm rms}\, 
  \Delta\eta\, m^2\, T\, \sum_{n=1}^\infty n^{-5/2}\, e^{n\mu_0 /T}\, 
  \bessk{2}\left ( \frac{nm}{T} \right)\, 
  \nonumber \\
  && \times 
  \left [ 1 + \frac{\eta_f}{2\sqrt{n}}\, e^{\eta_f^2/4n} 
    \int_{-\eta_f/2\sqrt{n}}^{\eta_f/2\sqrt{n}} e^{-x^2}\, 
    dx \right ] \, .
  \label{mbe6}
\eea
Note that the invariant volume cannot be factorized from the 
expression. This is generally true for models with a Bose-Einstein 
distribution with an $x$-dependent chemical potential. A special
case is the box-model where the invariant volume (\ref{box-vol}) 
still factorizes:
\beq
\label{mbe7}
\la N \ra = V_{\rm inv}^{\rm box}\,
\frac{m^2 T}{2\pi^2} \sum_{n=1}^{\infty}
n^{-3/2}\, e^{n\mu_0/T}\, \bessk{2}\left(\frac{nm}{T} \right ) .
\eeq
The different factorization properties of (\ref{mbe6}) and (\ref{mbe7})
imply that for $\eta_f = 0$ the yields are no longer related by a 
simple factor 2 once Bose statistics is taken into account.

Expressions (\ref{mbe6}) and (\ref{mbe7}) were used in the calculations 
of the direct pion multiplicity in Sec.~\ref{yield}. There, the chemical 
potential $\mu_0$ was set to 0 since chemical equilibrium was assumed.
The calculations in Sec.~\ref{psd-app-dist} were done with the
values of $\mu\equiv \mu_0$ given there.

%

%
\begin{table}
\begin{center}
\caption{Correlation radii for the Cartesian parametrization in the
LCMS obtained by calculating the averages of the squared radii from
the analysis of MTPC and VTPC data (see text). In the second row
the values of $K_\perp^{\rm bin}$ for each bin according to
(\ref{dan2})  are displayed. Selected rapidity bin: $3.9 < Y <
4.4$. (In fact, in the MTPC analysis the rapidity bin $4<Y<4.5$ was
used. Here that from VTPC data sets has been adopted.) Momenta are
given in units of GeV/$c$. In the \protect{$K_\perp$}-bin 
\protect{$0.3 - 0.45$}, data from \protect\cite{Stefan} for 
$0.3\, \mbox{GeV}/c <K_\perp<0.4\, \GeV/c$ and from \protect\cite{Harry} 
for $0.3\, \mbox{GeV}/c <K_\perp<0.45\, \GeV/c$ 
were averaged.}
\renewcommand{\arraystretch}{1.15}
\begin{tabular}{ccccc}
$K_\perp$-bin & 0 -- 0.1 & 0.1 -- 0.2 & 0.2 -- 0.3 & 0.3 -- 0.45${}$ \\
\hline
$K_\perp^{\rm bin}$& 0.065 & 0.151 &  0.247 & 0.365 \\ \hline
$R_s^2$ (fm${}^2$) &  $33.3 \pm 4.0$ & $27.0\pm 3.5$ & $25.0\pm 1.8$ & 
$21.4\pm 2.5$ \\
$R_o^2$ (fm${}^2$) & $38.7\pm 2.1$ & $38.8\pm 3.3$ & $33.6\pm 4.5$ &
$30.4\pm 5.2$ \\
$R_l^2$ (fm${}^2$) & $61.9\pm 5.3$ & $46.7 \pm 5.0$ & $27.4\pm 5.6$ &
$20.7\pm 3.1$ \\
$R_{ol}^2$ (fm${}^2$) & $8.4\pm 3.0$ & $13.8\pm 2.2$ & $8.4\pm 3.7$ &
$7.2\pm 5.6$ \\
\end{tabular}
\renewcommand{\arraystretch}{1}
\end{center}
\label{danal-bpdata}
\end{table}
%
\begin{table}
\begin{center}
\caption{The same as Table~\ref{danal-bpdata}, but for the 
correlation parameters in the Yano-Koonin-Pod\-go\-rets\-ki\u\i\
pa\-ra\-me\-tri\-za\-tion. $v_{_{\rm YK}}$ is measured relative to 
the LCMS. The same comments about binning as in 
Table~\ref{danal-bpdata} apply.}
\renewcommand{\arraystretch}{1.15}
\begin{tabular}{ccccc}
$K_\perp$-bin & 0 -- 0.1 & 0.1 -- 0.2 & 0.2 -- 0.3 & 0.3 -- 0.45${}$ \\
\hline
$K_\perp^{\rm bin}$& 0.065 & 0.151 &  0.247 & 0.365 \\ \hline
$R_s^2$ (fm${}^2$) &  $32.7 \pm 4.7$ & $28.4\pm 3.3$ & $26.0\pm 3.6$ & 
$23.2\pm 5.5$ \\
$R_\parallel^2$ (fm${}^2$) & $59.6\pm 9.8$ & $42.3\pm 5.8$ & $25.9\pm 6.7$ &
$21.1\pm 3.3$ \\
$R_0^2$ (fm${}^2$) & $28.6\pm 18.7$ & $14.8 \pm 9.2$ & $8.4\pm 6.7$ &
$8.1\pm 5.7$ \\
$v_{_{\rm YK}}$ & -$.15\pm .04$ & -$.26\pm .04$ & -$.23\pm .05$ &
-$.25\pm .10$ \\
\end{tabular}
\renewcommand{\arraystretch}{1}
\end{center}
\label{danal-ykpdata}
\end{table}
%
%
\begin{table}
\vspace*{-0.3cm}
\begin{center}
\renewcommand{\arraystretch}{1.15}
\caption{Sets of model parameters for box-shaped ({\sf b}{\tt *}) 
and Gaussian ({\sf g}{\tt *}) transverse density profiles, as obtained 
from fits to the single-particle $m_\perp$-spectrum and two-particle 
correlation radii. For each parameter set, the average transverse 
expansion velocity $\bar v_\perp$ and the total $\pi^-$ multiplicity 
$N$ (cf. Fig.~\ref{f-mult}) was calculated from the model. The errors 
for $N$ include only the uncertainties in the values of the model 
parameters.}
\vspace*{0.2cm}
\begin{tabular}{cccc}
Box {set} & {\sf b1} & {\sf b2} & {\sf b3} \\
\hline
$T$ (MeV) & 100 & 120 & 160 \\
$\eta_f$ & 0.6 & 0.5 & 0.35 \\
$R_B$ (fm) & $12.1\pm 0.2$ & $11.5\pm 0.2$ & $10.7\pm 0.2$ \\
$\tau_0$ (fm/$c$) & $6.3\pm 1.1$ & $5.5\pm 1.1$ & $4.4\pm 3.5$ \\
$\Delta\tau$ (fm/$c$) & $3.6\pm 0.6$ & $3.2\pm 0.7$ & $2.6 \pm 2.0$ \\
$\Delta\eta$ (fixed) & 1.3 & 1.3 & 1.3 \\
\hline $\bar v_\perp$ & 0.5 & 0.43 & 0.33 \\
$N$ & $139\pm 24$ & $254\pm 56$ & $813\pm 649$ \\ 
\hline\hline\hline
Gauss {set} & {\sf g1} & {\sf g2} & {\sf g3}  \\
\hline
$T$ (MeV) & 100 & 120 & 160 \\
$\eta_f$ & 0.6 & 0.48 & 0.35 \\
$R_G$ (fm) & $6.5 \pm 0.1$ &$5.9 \pm 0.2$ & $5.6 \pm 0.1$  \\
$\tau_0$ (fm/$c$) & $7.8 \pm 0.8$ & $6.6 \pm 0.9$ & $5.5 \pm 0.9$  \\
$\Delta\tau$ (fm/$c$) & $2.3 \pm 0.7$ & $2.3 \pm 0.7$ & $ 1.8 \pm 0.8$ \\
$\Delta\eta$ (fixed) & 1.3 & 1.3 & 1.3 \\
\hline
$\bar v_\perp$ & 0.46 & 0.39 & 0.29 \\
$N$ & $96 \pm 10$ & $157\pm 24$ & $544\pm 91$ 
\end{tabular}
\label{danal-sets}
\renewcommand{\arraystretch}{1}
\end{center}
\end{table} 
%
%
\begin{table}
\caption{The values of the effective intercept parameter  
as a function of $K_\perp$,  estimated from the NA49 data.}
\begin{center}
\begin{tabular}{ccccc}
$K_\perp$ (GeV/$c$) & 0.065 & 0.151 & 0.247 & 0.365 \\ 
$\lambda_{\rm dir}$ & 0.4 & 0.495 & 0.46 & 0.4 
\end{tabular}
\end{center}
\label{psd-t1}
\end{table}
%
%
\begin{table}
\caption{Estimated chemical potentials needed for the box-models of Table
        \ref{danal-sets} to fit the data for the average phase-space
        density.}
\begin{center}
\begin{tabular}{ccc}
set & $T$ (MeV) & $\mu$ (MeV)  \\
\hline
{\sf b1} & 100 & $123\pm 9$ \\
{\sf b2} & 120 & $115\pm15$ \\
{\sf b3} & 160 &  $90\pm 20$  
\end{tabular}
\end{center}
\label{psd-t2}
\end{table}
%
%
\begin{table}[b]
\begin{center}
\caption{Decay channels included in the calculation of the $K^-$ 
    contribution to the $h^-$-spectrum. Branching ratios were
    multiplied by the Clebsch-Gordan coefficients corresponding to
    the given isospin states.
    \label{comps-K}}
\vspace*{0.2cm}
\begin{tabular}{ccccc}
channel & $M_{\rm reson}\, (\MeV)$ & $\Gamma\, (\MeV)$ & $J$ & BR\\
\hline
$K^{*-} \to \pi^0 K^-$ & 892 & 115 & 1 & $1/3 \times 1$ \\
$\bar K^{*0} \to \pi^+ K^-$ & 892 & 115 & 1 & $2/3 \times 1$ 
\end{tabular}
\end{center} 
\vspace*{-0.4cm}
\end{table}
%
\begin{table}[b]
\vspace*{-.8cm}
\begin{center}
\caption{Decays contributing to the production of antiprotons. Branching
   ratios were multiplied by the appropriate Clebsch-Gordan coefficients.
   \label{comps-p}}
\vspace*{0.2cm}
\begin{tabular}{ccccc}
channel & $M_{\rm reson}\, (\MeV)$ & $\Gamma\, (\MeV)$ & $J$ & BR \\
\hline
$\bar \Delta^{++} \to \pi^- \bar p$ & 1232 & 115 & 3/2 & 1 \\
$\bar \Delta^+ \to \pi^0 \bar p$ & 1232 & 115 & 3/2 & $2/3 \times 1$ \\
$\bar \Delta^0 \to \pi^+ \bar p$ & 1232 & 115 & 3/2 & $1/3 \times 1$ \\[2ex]

$\bar \Lambda \to \pi^+ \bar p$ & 1116 & $\approx 0$& 1/2 & 0.639 \\
$\bar \Sigma^+ \to \pi^0 \bar p$ & 1193 & $\approx 0$& 1/2 & 0.516 \\
$\bar \Sigma^0 \to \bar \Lambda \gamma \to \pi^+ \bar p$ & 1193 
               & $\approx 0$& 1/2 & 0.639
\end{tabular}
\end{center} 
\vspace*{-0.4cm}
\end{table}
%

%
%
%
%
%
%

%
%
%

%

%

%
%

\end{document}